\documentclass[12pt]{iopart}
\usepackage[pdftex]{graphicx}
\expandafter\let\csname equation*\endcsname\relax
\expandafter\let\csname endequation*\endcsname\relax
\usepackage{amssymb}
\usepackage{amsmath}
\usepackage{aeitensor}
\usepackage{multirow}
\usepackage{url}
\usepackage{color}
\usepackage{textcomp}
\usepackage{hyperref}
\usepackage{acronym}
\newcommand{\appref}[1]{\ref{#1}}
\newcommand{\secref}[1]{section~\ref{#1}}
\newcommand{\figref}[1]{figure~\ref{#1}}

\newcommand{\Secref}[1]{Section~\ref{#1}}
\newcommand{\Figref}[1]{Figure~\ref{#1}}

\newcommand{\avec}{\vec{a}}
\newcommand{\bvec}{\vec{b}}
\newcommand{\cvec}{\vec{c}}
\newcommand{\dvec}{\vec{d}}

\newcommand{\ihat}{\vec{\imath}}
\newcommand{\jhat}{\vec{\jmath}}
\newcommand{\khat}{\vec{k}}

\newcommand{\cost}{\cos 2\psi}
\newcommand{\sint}{\sin 2\psi}

\newcommand{\Rx}{\A^{\onedot}}
\newcommand{\Ry}{\A^{\twodot}}
\newcommand{\Lx}{\A^{\tredot}}
\newcommand{\Ly}{\A^{\fordot}}

\newcommand{\Left}{{\textsc{l}}}
\newcommand{\Right}{{\textsc{r}}}
\newcommand{\AR}{A_\Right}
\newcommand{\AL}{A_\Left}
\newcommand{\tR}{\phi_\Right}
\newcommand{\tL}{\phi_\Left}

\newcommand{\MLEAR}{\MLE{A}_\Right}
\newcommand{\MLEAL}{\MLE{A}_\Left}
\newcommand{\MLEtR}{\MLE{\phi}_\Right}
\newcommand{\MLEtL}{\MLE{\phi}_\Left}
\newcommand{\LambdaR}{\Lambda_\Right}
\newcommand{\LambdaL}{\Lambda_\Left}

\newcommand{\mc}[1]{\mathcal{#1}}

\newcommand{\ep}{\tens{e}_{\!+}}
\newcommand{\ec}{\tens{e}_{\!\times}}
\newcommand{\epsp}{\tens{\varepsilon}_{\!+}}
\newcommand{\epsc}{\tens{\varepsilon}_{\!\times}}

\newcommand{\cosi}{\chi}
\newcommand{\A}{\mc{A}}
\newcommand{\MLE}[1]{\widehat{#1}}

\newcommand{\M}{\mc{M}}
\newcommand{\circidx}[1]{\breve{#1}}
\newcommand{\mudot}{\circidx{\mu}}
\newcommand{\nudot}{\circidx{\nu}}

\newcommand{\onedot}{\circidx{1}}
\newcommand{\twodot}{\circidx{2}}
\newcommand{\tredot}{\circidx{3}}
\newcommand{\fordot}{\circidx{4}}

\newcommand{\F}{\mc{F}}
\newcommand{\B}{\mc{B}}
\newcommand{\Hs}{\mc{H}_s}
\newcommand{\Hn}{\mc{H}_n}
\newcommand{\Hf}{\mc{H}_f}

\newcommand{\abs}[1]{\left\lvert#1\right\rvert}

\newcommand{\tens}[1]{\aeitensor{#1}}
\newcommand{\un}[1]{\text{\,#1}}
\newcommand{\cft}[1]{\widetilde{#1}}
\newcommand{\tssb}{\tau}

\DeclareMathOperator{\Real}{Re}

\newcommand{\pdf}{\text{pdf}}
\newcommand{\Tsft}{T_{\text{sft}}}
\newcommand{\Tobs}{T_{\text{obs}}}

\newcommand{\coord}{coordinate}

\newcommand{\coeff}{coefficient}

\newcommand{\iSFT}{l}
\newcommand{\sumXiSFT}{\sum_{X\iSFT}}
\newcommand{\scalar}[2]{\left(#1|#2\right)}
\newcommand{\detV}[1]{{#1}}

\newcommand{\dcc}{LIGO-P1800060-v5}

\numberwithin{equation}{section}

\def\commitDATE{ Sat Aug 4 21:20:29 2018 -0400}

 \newcommand{\PKA}{0.154}
\newcommand{\PKB}{0.234}
\newcommand{\PKC}{-0.0104}
\newcommand{\PKI}{0.388}
\newcommand{\PKK}{-0.0207}
\newcommand{\PKL}{-0.0805}
\newcommand{\PKKI}{-0.0533}
\newcommand{\PKLI}{-0.207}

\newcommand{\OiI}{0.373}
\newcommand{\OiK}{-0.0120}
\newcommand{\OiL}{-0.0385}
\newcommand{\OiKI}{-0.0321}
\newcommand{\OiLI}{-0.103}

\newcommand{\oI}{0.679}
\newcommand{\oK}{0.1604}
\newcommand{\oL}{0.6527}
\newcommand{\oKI}{0.236}
\newcommand{\oLI}{0.961}

\newcommand{\HiI}{0.305}

\newcommand{\HiL}{-0.1479}
\newcommand{\HiKI}{0.0000}
\newcommand{\HiLI}{-0.485}
 
\begin{document}
\title[Analytic $\B$-stat Approximation]
{An Analytic Approximation\\ to the Bayesian Detection Statistic\\
  for Continuous Gravitational Waves
}
\author{John J.\ Bero}
\address{Center for Computational Relativity and Gravitation
  and School of Physics and Astronomy, Rochester Institute of Technology,
  84 Lomb Memorial Drive, Rochester, NY 14623, USA}
\author{John T.\ Whelan}
\ead{john.whelan@astro.rit.edu}
\address{Center for Computational Relativity and Gravitation
  and School of Mathematical Sciences, Rochester Institute of Technology,
  85 Lomb Memorial Drive, Rochester, NY 14623, USA}
\date{\commitDATE
}
\begin{abstract}
  We consider the Bayesian detection statistic for a targeted search
  for continuous gravitational waves, known as the $\B$-statistic.
  This is a Bayes factor between signal and noise hypotheses, produced
  by marginalizing over the four amplitude parameters of the signal.
  We show that by Taylor-expanding to first order in certain averaged
  combinations of antenna patterns (elements of the parameter space
  metric), the marginalization integral can be performed analytically,
  producing a closed-form approximation in terms of confluent
  hypergeometric functions.  We demonstrate using Monte Carlo
  simulations that this approximation is as powerful as the full
  $\B$-statistic, and outperforms the traditional maximum-likelihood
  $\F$-statistic, for several observing scenarios which involve an
  average over sidereal times.  We also show that the approximation does
  not perform well for a near-instantaneous observation, so the
  approximation is suited to long-time continuous wave observations
  rather than transient modelled signals such as compact binary
  inspiral.
\end{abstract}
\maketitle

\acrodef{NS}[NS]{neutron star}
\acrodef{GW}[GW]{gravitational wave}
\acrodef{SSB}[SSB]{solar-system barycenter}
\acrodef{JKS}[JKS]{Jaranowski-Kr\'{o}lak-Schutz}

\section{Introduction}

The signal from a non-precessing source of \acp{GW} such as a rotating
neutron star or slowly-evolving binary system, can be described by
phase-modulation parameters, which determine the shape of the signal,
and amplitude parameters.  In the case where the phase-modulation
parameters are assumed to be known, the likelihood ratio between
models with and without signal is a function of the four amplitude
parameters.  Jaranowski Kr\'{o}lak and Schutz \cite{JKS}
constructed a maximum-likelihood statistic (known as the
$\F$-statistic), which is the basis of many existing searches for
continuous \acp{GW}.  Prix and Krishnan \cite{PK} proposed
a Bayesian alternative (the $\B$-statistic) which instead marginalized
the likelihood ratio over these parameteres, assuming a
geometrically-inspired prior distribution.  Exact evaluation of the
$\B$-statistic requires integration over the four-dimensional
amplitude parameter space; Whelan et al \cite{WPCW} showed
that two of the integrals can be done analytically, but a
two-dimensional numerical integration remains.  They also showed that
the marginalization integrals can be done exactly if the
parameter-space metric (determined by averaged combinations of antenna
patterns) has a block-diagonal form.  In this paper, we generalize
this result to produce an analytical approximation to the
$\B$-statistic by Taylor expanding to first order in the off-diagonal
metric elements.

This paper is laid out as follows: in \secref{s:formalism} we give a
brief overview of the background information and formalism related to
this topic, including a discussion of GW signal analysis and a
description of the two detection statistics which already exist.
\Secref{s:B-stat_approx} contains the derivation of our approximation
and in \secref{s:evaluation} we test the power of the approximation as
a detection statistic.  \Secref{s:conclusions} concludes with a
summary of the results and their practical implications.

\section{Formalism}
\label{s:formalism}

\subsection{Signal Parameters}

We follow the conventions and notation of \cite{WPCW}, where
more details can be found.  We summarize the relevant expressions
here.  For a \ac{GW} signal coming from a sky position specified by
right ascension $\alpha$ and declination $\delta$, we can define a
propagation unit vector $\khat$ pointing from the source to the
\ac{SSB}.  The tensor \ac{GW} can then be resolved in a basis of
traceless tensors transverse to $\khat$ as
\begin{equation}
  \label{e:signal}
  \tens{h}(\tssb) = h_+(\tssb)\, \ep + h_\times(\tssb) \, \ec
  \ .
\end{equation}
For a nearly periodic signal, as from a rotating \ac{NS}, the
polarization components are
\begin{equation}
  \label{eq:signal2}
  h_+(\tssb) \equiv \frac{h_0}{2} (1+\cosi^2) \cos [\phi(\tssb)+\phi_0]
  \qquad\hbox{and}\qquad
  h_\times(\tssb) \equiv h_0 \cosi \sin [\phi(\tssb)+\phi_0]
  \ ,
\end{equation}
where $\cosi=\cos\iota$ is the cosine of angle between the line of
sight and the neutron star's rotation axis, and
\begin{equation}
  h_0 = \frac{4G}{c^4}\frac{\abs{I_{xx}-I_{yy}}\Omega^2}{d}
\end{equation}
is the amplitude in terms of the equatorial quadrupole moments
$\{I_{xx},I_{yy}\}$, the rotation frequency $\Omega$, and the distance
$d$ to the source.  The preferred polarization basis tensors are given
by
\begin{equation}
  \label{e:tensorrot}
  \ep\, = \epsp\,\cost\,
  +\,\epsc\,\sint
  \qquad\hbox{and}\qquad
  \ec\, = -\,\epsp\,\sint\,
  +\,\epsc\,\cost
  \ .
\end{equation}
where $\epsp = \ihat\otimes\ihat - \jhat\otimes\jhat$ and $\epsc =
\ihat\otimes\jhat + \jhat\otimes\ihat$ are the fiducial basis tensors
defined using unit vectors orthogonal to $\khat$, with $\ihat$
pointing ``West on the sky'' in the direction of decreasing right
ascension $\alpha$, and $\jhat$ pointing ``North on the sky'' in the
direction of increasing declination $\delta$.  The polarization angle
$\psi$ measures the angle counter-clockwise on the sky from $\ihat$ to
the \ac{NS}'s equatorial plane.

The phase evolution $\phi(\tssb)$ in terms of the arrival time $\tssb$
at the \ac{SSB} can be written in terms of \ac{NS} rotation or
spindown parameters, e.g.,
\begin{equation}
  \phi(\tssb) = 2\pi \left(f_0\tssb + f_1\frac{\tssb^2}{2} + \cdots\right)
  \ ,
\end{equation}
although it may be more complicated, e.g., for \acp{NS} in binary
systems.

The strain, $h$, measured by an interferometric GW detector whose arms
are parallel to the unit vectors $\vec{p}_1$ and $\vec{p}_2$ is given
by
\begin{equation}
\label{eq:strain}
h = \tens{h}:\tens{d}
\end{equation}
where\footnote{We limit attention in this section to the
  long-wavelength limit, where the detectors are assumed to be small
  compared to the gravitational wavelength $c/f_0$, which is
  appropriate to most observations with ground-based interferometric
  detectors.  At higher frequencies, the detector tensor $\tens{d}(f)$
  is frequency-dependent and complex.  See e.g.,\cite{WPCW}
  for more details.}
\begin{equation}
\tens{d} = \frac{\vec{p}_1\otimes\vec{p}_1-\vec{p}_2\otimes\vec{p}_2}{2}
\end{equation}
is the detector tensor and $:$ signifies the double dot product,
defined by
$(\avec\otimes\bvec):(\cvec\otimes\dvec)=(\avec\cdot\dvec)(\bvec\cdot\cvec)$.
The GW strain can also be expressed as
\begin{equation}
h = h_+F_+ + h_\times F_\times
\ ,
\end{equation}
where $F_+$ and $F_\times$ are the detector antenna pattern functions
which depend on the 3 angles defining the source sky position and
polarization basis relative to your detector, which in our case would
be the right ascension $\alpha$, the declination $\delta$ and the
polarization angle $\psi$. If we separate out their dependence on
$\psi$, then the pattern functions have the form
\begin{subequations}
  \begin{alignat}{3}
    F_+(\alpha,\delta,\psi)\, &= &a(\alpha,\delta)&\,\cos2\psi\,
    &+&\,b(\alpha,\delta)\,\sin2\psi
\ 
    \\
    F_\times(\alpha,\delta,\psi)\, &= -\,&a(\alpha,\delta)&\,\sin2\psi\,
    &+&\,b(\alpha,\delta)\,\cos2\psi
\ ,
  \end{alignat}
\end{subequations} 
where $a$ and $b$ are amplitude modulation coefficients defined in
terms of the detector tensor $\tens{d}$ as
\begin{subequations}
\label{eq:AM_coeff}
  \begin{align}
  a &\equiv \epsp:\tens{d}
\ ,\\
  b &\equiv \epsc:\tens{d}
\ .
  \end{align}
\label{eq:def-a-b}
\end{subequations}
These coefficients are defined with respect to the reference
polarization basis and depend both on the sky position of the GW
source and the sidereal time at which the measurement is taking place.

It is useful to divide the signal parameters into \textit{amplitude
  parameters} $\{h_0,\cosi,\psi,\phi_0\}$ and \textit{phase-evolution
  parameters} such as the sky position $\{\alpha,\delta\}$, and any
parameters describing $\phi(\tssb)$.  The dependence of the signal on
the amplitude parameters can be written simply
as\cite{JKS,WPCW}
\begin{equation}
  \label{e:tenswavedot}
  \tens{h}(\tssb;\A,\lambda) = \A^{\mudot} \,\tens{h}_{\mudot}(\tssb;\lambda)
  \ ,
\end{equation}
where the Einstein summation convention implies the sum
$\sum_{\mu=1}^4$ over repeated indices.  The 
amplitudes $\{\A^{\mudot}\}$ are defined by\footnote{Our {\coord}s
  $\{\A^{\mudot}\}$, introduced in \cite{WPCW}, are related
  to the more familiar \ac{JKS} {\coord}s $\{\A^{\mu}\}$ of
  \cite{JKS} by $\A^1=\Rx+\Lx$, $\A^2=\Ry-\Ly$,
  $\A^3=-\Ry-\Ly$, $\A^4=\Rx-\Lx$.}
\begin{subequations}
  \label{e:PQpolar}
  \begin{align}
    \Rx = \AR\cos\tR
    \qquad&\hbox{and}\qquad
    \Ry = \AR\sin\tR
    \\
    \Lx = \AL\cos\tL
    \qquad&\hbox{and}\qquad
    \Ly = \AL\sin\tL
    \ ;
  \end{align}
\end{subequations}
where
\begin{subequations}
  \label{e:pqphys}
  \begin{align}
    \AR = h_0\left(\frac{1+\cosi}{2}\right)^2
    \quad&\hbox{and}\quad
    \tR = \phi_0+2\psi
    \ ;
    \\
    \AL = h_0\left(\frac{1-\cosi}{2}\right)^2
    \quad&\hbox{and}\quad
    \tL = \phi_0-2\psi
  \end{align}
\end{subequations}
are the amplitudes and phases of the right- and
left-circularly-polarized components of the signal, respectively.

\subsection{Likelihood Function and Detection Statistics}

If we denote the data recorded in the \ac{GW} detector(s) as
$\detV{x}$, and assume those data to consist of the signal
$\A^{\mudot}\detV{h}_{\mudot}$ plus Gaussian noise, the sampling
distribution for the data will be
\begin{equation}
  \pdf(\detV{x}|\A) \propto
  \exp\left(
    -\frac{1}{2}\scalar{\detV{x}-\A^{\mudot}\detV{h}_{\mudot}}
    {\detV{x}-\A^{\mudot}\detV{h}_{\mudot}}
  \right)
\end{equation}
The log-likelihood ratio will thus be
\begin{equation}
  \label{e:loglikeAx}
  \Lambda(\{\A^{\mudot}\};\detV{x})
  = \ln \frac{\pdf(\detV{x}|\A)}{\pdf(\detV{x}|0)}
  = \A^{\mudot} x_{\mudot}
  - \frac{1}{2}\A^{\mudot} \M_{\mudot\nudot} \A^{\nudot}
\end{equation}
where $x_{\mudot} \equiv \scalar{\detV{x}}{\detV{h}_{\mudot}}$ is the
scalar product (see \ref{app:metric}) of the data with the template
waveform, and
\begin{equation}
  \{\M_{\mudot\nudot}\} \equiv \{\scalar{\detV{h}_{\mudot}}{\detV{h}_{\nudot}}\}
  =
  \begin{pmatrix}
    I &  0 &  L & -K \\
    0 &  I &  K &  L \\
    L &  K &  J &  0 \\
   -K &  L &  0 &  J
  \end{pmatrix}
  \label{eq:Mmunu}
\end{equation}
forms a metric on parameter space.

If we define $\{\M^{\mudot\nudot}\}$ as the matrix inverse of
$\{\M_{\mudot\nudot}\}$, we can write the maximum-likelihood values of
the amplitude parameters $\{\A^{\mudot}\}$ as
\begin{equation}
  \MLE{\A}^{\mudot}(\detV{x}) = \M^{\mudot\nudot} x_{\nudot}
  \ ,
\end{equation}
Since the maximum-likelihood parameters
$\{\MLE{\A}^{\mudot}(\detV{x})\}$ contain equivalent information to
the projections $\{x_{\nudot}\}$ (which form jointly sufficient
statistics for the amplitude parameters $\A$), we can use
$\{\MLE{\A}^{\mudot}\}$ as a representation of the relevant part of
the data.  Their sampling distribution can be written as the
multivariate Gaussian
\begin{equation}
  \pdf(\MLE{\A}|\A) =
  \left(
    \det 2\pi\M
  \right)^{-1/2}
  \exp\left(
    -\frac{1}{2}(\MLE{\A}^{\mudot}-\A^{\mudot})
    \M_{\mudot\nudot}
    (\MLE{\A}^{\nudot}-\A^{\nudot})
  \right)
\end{equation}
This is useful for conducting Monte Carlo simulations (as was done in
\cite{PK}): one need not simulate the full \ac{GW} data,
only generate draws of the four maximum-likelihood parameters
$\{\MLE{\A}^{\nudot}\}$ representing the data.

It is also convenient to write the log-likelihood ratio in terms of
$\MLE{\A}$ as well:
\begin{equation}
  \Lambda(\A;\MLE{\A})
  = \A^{\mudot} \M_{\mudot\nudot} \MLE{\A}^{\nudot}
  - \frac{1}{2}\A^{\mudot} \M_{\mudot\nudot} \A^{\nudot}
  \ .
\label{eq:likelihood}
\end{equation}
This is written explicitly in terms of the polar representation in
\appref{app:taylor}.

The $\F$-statistic\cite{JKS} is defined as the maximized
log-likelihood ratio,
\begin{equation}
  \label{e:Fstat}
  \begin{split}
    \F &= \max_{\A} \Lambda(\A;\detV{x})
    = \Lambda(\MLE{\A};\detV{x})
    = \frac{1}{2}\MLE{\A}^{\mudot} \M_{\mudot\nudot} \MLE{\A}^{\nudot}
    \\
    &= \frac{1}{2}I\MLEAR^2
    + \frac{1}{2}J\MLEAL^2
    + \MLEAR\MLEAL
    \left[
      K\sin(\MLEtR-\MLEtL) + L\cos(\MLEtR-\MLEtL)
    \right]
  \end{split}
\end{equation}
The $\B$-statistic\cite{PK} is defined as the Bayes factor
between models with and without signal:
\begin{equation}
  \B(\detV{x}) = \frac{\pdf(\detV{x}|\Hs)}{\pdf(\detV{x}|\Hn)}
  = \frac{\int\pdf(\detV{x}|\A)\,\pdf(\A|\Hs)\,d^4\!\A}{\pdf(\detV{x}|0)}
  = \int e^{\Lambda(\{\A^{\mudot}\};\detV{x})}\,\pdf(\A|\Hs)\,d^4\!\A
\end{equation}
The prior is taken to be uniform in $\cosi\in(-1,1)$, $\psi\in(-\pi/4.\pi/4)$
and $\phi_0\in(0,2\pi)$, so that
\begin{equation}
  \label{e:Bstatprior}
  \pdf(h_0,\cosi,\psi,\phi_0|\Hs) = \frac{\pdf(h_0|\Hs)}{2\pi^2}
\end{equation}
The convention introduced in \cite{PK} is to use an
improper prior $\pdf(h_0|\Hs)=A$, $0<h<\infty$, so that
\begin{equation}
  \label{e:Bstatexplicit}
  \begin{split}
    \B(\detV{x})
    &=
    \frac{A}{2\pi^2}
    \int_{0}^{2\pi}
    \int_{-1}^{1}
    \int_{-\pi/4}^{\pi/4}
    \int_{0}^{\infty}
    e^{\Lambda(\{\A^{\mudot}\};\detV{x})}
    \,dh_0
    \,d\psi
    \,d\cosi
    \,d\phi_0
    \\
    &=
    \frac{A}{8\pi^2}
    \int_{0}^{2\pi}
    \int_{0}^{2\pi}
    \int_{0}^{\infty}
    \int_{0}^{\infty}
    e^{\Lambda(\{\A^{\mudot}\};\detV{x})}
    \frac{d\AR\,d\AL\,d\tR\,d\tL}{\sqrt{\AR\AL}}
  \end{split}
\end{equation}

\section{An approximate form for the $\B$-statistic}
\label{s:B-stat_approx}

Previous work \cite{WPCW} showed that the $\B$-statistic
integral \eqref{e:Bstatexplicit} can be exactly evaluated in the case
where $K=0=L$, so that the metric \eqref{eq:Mmunu} becomes diagonal
and the left- and right-circularly polarized subspaces decouple.  We
show in \appref{app:metric} that $K$ and $L$ can be small compared to
$I=J$, especially in continuous-wave observations containing an
average over sidereal times and/or detectors.

When $K$ and $L$ are small compared to $I$ and $J$, it is fruitful to
consider a Taylor expansion of the B-statistic integral
\eqref{e:Bstatexplicit}, which we carry out in \appref{app:taylor},
and find
\begin{equation}
  \B(\detV{0}) = \frac{A[\Gamma(\frac{1}{4})]^2}{2^{5/2}(IJ)^{1/4}}
\end{equation}
and
\begin{multline}
  \label{e:BstatApprox}
  \ln\frac{\B(\detV{x})}{\B(\detV{0})}
  \approx
  \ln{}_1F_1\left(\frac{1}{4},1,\frac{I\MLEAR^2}{2}\right)
  + \ln{}_1F_1\left(\frac{1}{4},1,\frac{J\MLEAL^2}{2}\right)
  +
  \left[
    K\sin(\MLEtR-\MLEtL) + L\cos(\MLEtR-\MLEtL)
  \right]
  \MLEAR\MLEAL
  \vphantom{\frac{{}_1F_1\left(\frac{5}{4},2,\frac{J\MLEAL^2}{2}\right)}{{}_1F_1\left(\frac{1}{4},1,\frac{J\MLEAL^2}{2}\right)}}
  \\
  \times
  \left[
    \frac{1}{4}
    \left(
      \frac{{}_1F_1\left(\frac{5}{4},2,\frac{I\MLEAR^2}{2}\right)}{{}_1F_1\left(\frac{1}{4},1,\frac{I\MLEAR^2}{2}\right)}
    \right)
    + \frac{1}{4}
    \left(
      \frac{{}_1F_1\left(\frac{5}{4},2,\frac{J\MLEAL^2}{2}\right)}{{}_1F_1\left(\frac{1}{4},1,\frac{J\MLEAL^2}{2}\right)}
    \right)
    -
    \frac{1}{16}
    \left(
      \frac{{}_1F_1\left(\frac{5}{4},2,\frac{I\MLEAR^2}{2}\right)}{{}_1F_1\left(\frac{1}{4},1,\frac{I\MLEAR^2}{2}\right)}
    \right)
    \left(
      \frac{{}_1F_1\left(\frac{5}{4},2,\frac{J\MLEAL^2}{2}\right)}{{}_1F_1\left(\frac{1}{4},1,\frac{J\MLEAL^2}{2}\right)}
    \right)
  \right]
\end{multline}
where the terms omitted are second order and higher in $K$ and/or $L$.

We can compare this to several limiting cases and alternative forms.
First, note that if $K=0=L$, we recover the result of section~6.1 of
\cite{WPCW}.  [See equation~(6.11) of that work.]  Second, in
the limit that $\MLEAR$ and $\MLEAL$ are both large, the asymptotic
form of the confluent hypergeometric functions [see identity (13.5.1)
of \cite{abramowitz64:_handb_mathem_funct}]
\begin{subequations}
  \begin{align}
    {}_1F_1\left(\frac{1}{4},1,\frac{I\MLE{A}^2}{2}\right)
    &\stackrel{\MLE{A}\rightarrow\infty}{\longrightarrow}
    \frac{1}{\Gamma(\frac{1}{4})}
    \left(\frac{I\MLE{A}^2}{2}\right)^{\!-3/4}\! e^{I\MLE{A}^2/2}
    \\
    {}_1F_1\left(\frac{5}{4},2,\frac{I\MLE{A}^2}{2}\right)
    &\stackrel{\MLE{A}\rightarrow\infty}{\longrightarrow}
    \frac{1}{\Gamma(\frac{5}{4})}
    \left(\frac{I\MLE{A}^2}{2}\right)^{\!-3/4}\! e^{I\MLE{A}^2/2}
  \end{align}
\end{subequations}
says that
\begin{multline}
  \label{e:Bapproxasymptotic}
  \ln\frac{\B(\detV{x})}{\B(\detV{0})}
  \stackrel{\MLEAR,\MLEAL\rightarrow\infty}{\longrightarrow}
  \frac{1}{2}I\MLEAR^2
  + \frac{1}{2}J\MLEAL^2
  + \left[
    K\sin(\MLEtR-\MLEtL) + L\cos(\MLEtR-\MLEtL)
  \right]
  \MLEAR\MLEAL
  \\
  -\frac{3}{4}\ln\left(\frac{1}{2}I\MLEAR^2\right)
  -\frac{3}{4}\ln\left(\frac{1}{2}J\MLEAL^2\right)
  -2\ln\Gamma\left(\frac{1}{4}\right)
  \\
  = \F - \frac{3}{2}\ln(\MLEAR\MLEAL) + \text{const}
\end{multline}
which is the result in equation (5.37) of \cite{WPCW}.

\section{Evaluation of Approximation}
\label{s:evaluation}

We evaluate the approximation for three cases of interest, which are
further detailed in \appref{app:metric}:
\begin{enumerate}
\item The case originally considered in \cite{PK}: a
  $\Tobs=25\un{hr}$ observation of a source at right ascension
  $2\un{radians}$, declination $-0.5\un{radians}$, with a single detector
  (LIGO Hanford, known as H1) beginning at GPS time 756950413 (2014
  Jan 1 at 00:00 UTC), for which $I=J=\PKI\frac{\Tobs}{S_n(f_0)}$,
  $K=\PKK\frac{\Tobs}{S_n(f_0)}$, and
  $L=\PKL\frac{\Tobs}{S_n(f_0)}$. so $K/I=\PKKI$ and $L/I=\PKLI$.
  This is a typical long-observation case.\footnote{Note that this
    case is slightly less favorable than another realistic alternative
    with the same sky position, which averages over the O1 segments
    from LIGO Hanford and LIGO Livingston, for which
    $I=J=\OiI\frac{\Tobs}{S_n(f_0)}$, $K=\OiK\frac{\Tobs}{S_n(f_0)}$,
    and $L=\OiL\frac{\Tobs}{S_n(f_0)}$. so $K/I=\OiKI$ and
    $L/I=\OiLI$.  However, as we shall see, the approximation performs
    well enough for the case considered that this more favorable case
    would be a redundant illustration.}
\item An observation with perfect sidereal-time averaging of a source
  on the celestial equator (declination $0$) using only H1.  As shown
  in \appref{app:metric}, this is a worst-case long-observation
  scenario, for which $I=J=\HiI\frac{\Tobs}{S_n(f_0)}$,
  $K=0$, and
  $L=\HiL\frac{\Tobs}{S_n(f_0)}$. so $K/I=0$ and $L/I=\HiLI$.  It
  provides an intermediate case where the approximation has not broken
  down completely.
\item An short two-detector (LIGO Hanford and Livingston) observation
  of a source at right ascension $2\un{radians}$, declination
  $-0.5\un{radians}$, at Greenwich sidereal time 00:00, for which
  $I=J=\oI\frac{\Tobs}{S_n(f_0)}$, $K=\oK\frac{\Tobs}{S_n(f_0)}$, and
  $L=\oL\frac{\Tobs}{S_n(f_0)}$. so $K/I=\oKI$ and $L/I=\oLI$.  This
  is a case where we do not expect the approximation to perform well.
\end{enumerate}

\subsection{Numerical Evaluation of $\B$-statistic Integral}

To compare our approximate form of the $\B$-statistic to its exact
value, we have to evaluate the integral \eqref{e:Bstatexplicit}.  It
was shown in \cite{WPCW} that the log-likelihood ratio
\eqref{e:loglikeAx} can be written in physical {\coord}s as
\begin{equation}
  \Lambda(\{\A^{\mudot}\};\detV{x})
  = h_0\,\omega(\detV{x};\cosi,\psi)
  \cos(\phi_0-\varphi_0(\detV{x};\cosi,\psi))
  - \frac{h_0^2 [\gamma(\cosi,\psi)]^2}{2}
\end{equation}
and the $h_0$ and $\phi_0$ integrals performed explicitly to reduce
the $\B$-statistic to a double integral\footnote{A similar reduction
  to a two-dimensional integral appears in \cite{Dergachev2012},
  with the integrand empirically estimated rather than evaluated
  analytically.}
\begin{equation}
  \B(\detV{x}) = \frac{A}{\sqrt{2\pi}}
  \int_{-1}^{1}
  \int_{-\pi/4}^{\pi/4}
  \frac{I_0(\zeta(\detV{x};\cosi,\psi))\,e^{\zeta(\detV{x};\cosi,\psi)}}
  {\gamma(\cosi,\psi)}
  \,d\psi
  \,d\cosi
  \ ,
\end{equation}
where
\begin{equation}
  \zeta(\detV{x};\cosi,\psi)
  = \frac{[\omega(\detV{x};\cosi,\psi)]^2}{4[\gamma(\cosi,\psi)]^2}
  \ .
\end{equation}
We note here the explicit forms of $\gamma(\cosi,\psi)$ and
$\omega(\detV{x};\cosi,\psi)$.  (The form of
$\varphi_0(\detV{x};\cosi,\psi)$ is irrelevant to the result of the
integral.)  From \eqref{e:AMA} we can see
\begin{equation}
  \label{eq:gamma}
  \begin{split}
    \gamma(\cosi,\psi)^2
    &=
    \frac{\A^{\mudot} \M_{\mudot\nudot} \A^{\nudot}}{h_0^2}
    =
    \frac{1}{h_0^2}\left[I\AR^2 + J\AL^2
      + 2\AR\AL\left[K\sin(\tR-\tL) + L\cos(\tR-\tL)\right]\right]
    \\
    &= I\left(\frac{1+\cosi}{2}\right)^4 + J\left(\frac{1-\cosi}{2}\right)^4
    + 2\left(\frac{1+\cosi}{2}\right)^2\left(\frac{1-\cosi}{2}\right)^2
    \left[K\sin(4\psi) + L\cos(4\psi)\right]
    \ .
  \end{split}
\end{equation}
while
\begin{multline}
  \omega(\detV{x};\cosi,\psi)
  \cos(\phi_0-\varphi_0(\detV{x};\cosi,\psi))
  = \frac{\A^{\mudot}x_{\mudot}}{h_0}
  \\
  = \left(\frac{1+\cosi}{2}\right)^2 (x_{\onedot}\cos\tR+x_{\twodot}\sin\tR)
  + \left(\frac{1-\cosi}{2}\right)^2 (x_{\tredot}\cos\tL+x_{\fordot}\sin\tL)
  \\
  = U\cos\phi_0 + V\sin\phi_0
\end{multline}
so $[\omega(\detV{x};\cosi,\psi)]^2=U^2+V^2$, where
\begin{subequations}
  \begin{align}
    U
    &=
      \cos 2\psi
      \left[
      \left(\frac{1+\cosi}{2}\right)^2 x_{\onedot}
      + \left(\frac{1-\cosi}{2}\right)^2 x_{\tredot}
      \right]
      + \sin 2\psi
      \left[
      \left(\frac{1+\cosi}{2}\right)^2 x_{\twodot}
      - \left(\frac{1-\cosi}{2}\right)^2 x_{\fordot}
      \right]
    \\
    V
    &=
      \cos 2\psi
      \left[
      \left(\frac{1+\cosi}{2}\right)^2 x_{\twodot}
      + \left(\frac{1-\cosi}{2}\right)^2 x_{\fordot}
      \right]
      - \sin 2\psi
      \left[
      \left(\frac{1+\cosi}{2}\right)^2 x_{\onedot}
      - \left(\frac{1-\cosi}{2}\right)^2 x_{\tredot}
      \right]
  \end{align}
\end{subequations}
The simulations that follow, we evaluate the integrals for the
$\B$-statistic using a 3000-point Monte Carlo integration on the space
$\cosi\in(-1,1)$, $\psi\in(-\pi/4,\pi/4)$.  This has the advantage
that, even when the integrand depends only weakly on $\psi$, we still
estimate the $\cosi$ integral accurately.

\subsection{Comparison of Statistic Values}

We compare our approximation to the numerically-evaluated exact
$\B$-statistic, and to the $\F$-statistic.  Each statistic is a
function of the four data values $\{x_{\mudot}\}$.  However, if we
express it in terms of the maximum-likelihood parameters
$\{\MLE{\A}^{\mudot}\}$, we see that all of the statistics are
independent of the combination $\MLEtR+\MLEtL=2\MLE{\phi}_0$ and
depend on the angles $\MLEtR$ and $\MLEtL$ only in the combination
$\MLEtR-\MLEtL=4\MLE{\psi}$.  Thus we can consider the statistics on
the three-dimensional space parameterized by $\MLEAR\ge 0$, $\MLEAL\ge
0$, and $\MLEtR-\MLEtL\in[0,2\pi)$.  For visualization purposes, we
plot contours of constant statistic versus $\MLEAR$ and $\MLEAL$ on
slices of constant $\MLEtR-\MLEtL$, in analogy to Figure 3 of
\cite{WPCW}, which considered a metric with $K=0=L$, for
which the statistics were independent of $\MLEtR$ and $\MLEtL$.  If we
plot $\MLEtR-\MLEtL=0$ in the first quadrant and $\MLEtR-\MLEtL=\pi$
in the second, we are effectively plotting $\MLEAR\cos(\MLEtR-\MLEtL)$
versus $\MLEAL$ on the slice $\sin(\MLEtR-\MLEtL)=0$.  Likewise, if we
plot $\MLEtR-\MLEtL=\frac{\pi}{2}$ in the first quadrant and
$\MLEtR-\MLEtL=-\frac{\pi}{2}$ in the second, we are effectively
plotting $\MLEAR\sin(\MLEtR-\MLEtL)$ versus $\MLEAL$ on the slice
$\cos(\MLEtR-\MLEtL)$.  Since the approximate $\B$-statistic and the
$\F$-statistic both depend on the combination $K\sin(\tR-\tL) +
L\cos(\tR-\tL)$, the former slice focuses on the impact of $L$ and the
second on the impact of $K$.  Note that another choice of slice would
be to chose $\tR-\tL=\tan^{-1}\left(-\frac{L}{K}\right)$, so that the
$K$-and-$L$-dependent part of the statistics vanished, or
$\tR-\tL=\tan^{-1}\left(\frac{K}{L}\right)$, which would maximize the
impact of this term.  In practice, for the examples we chose,
$\abs{L}$ is significantly larger than $\abs{K}$, so these slices
would be similar to the ones we plot.

\begin{figure}
\begin{center}
\includegraphics[width=\textwidth]{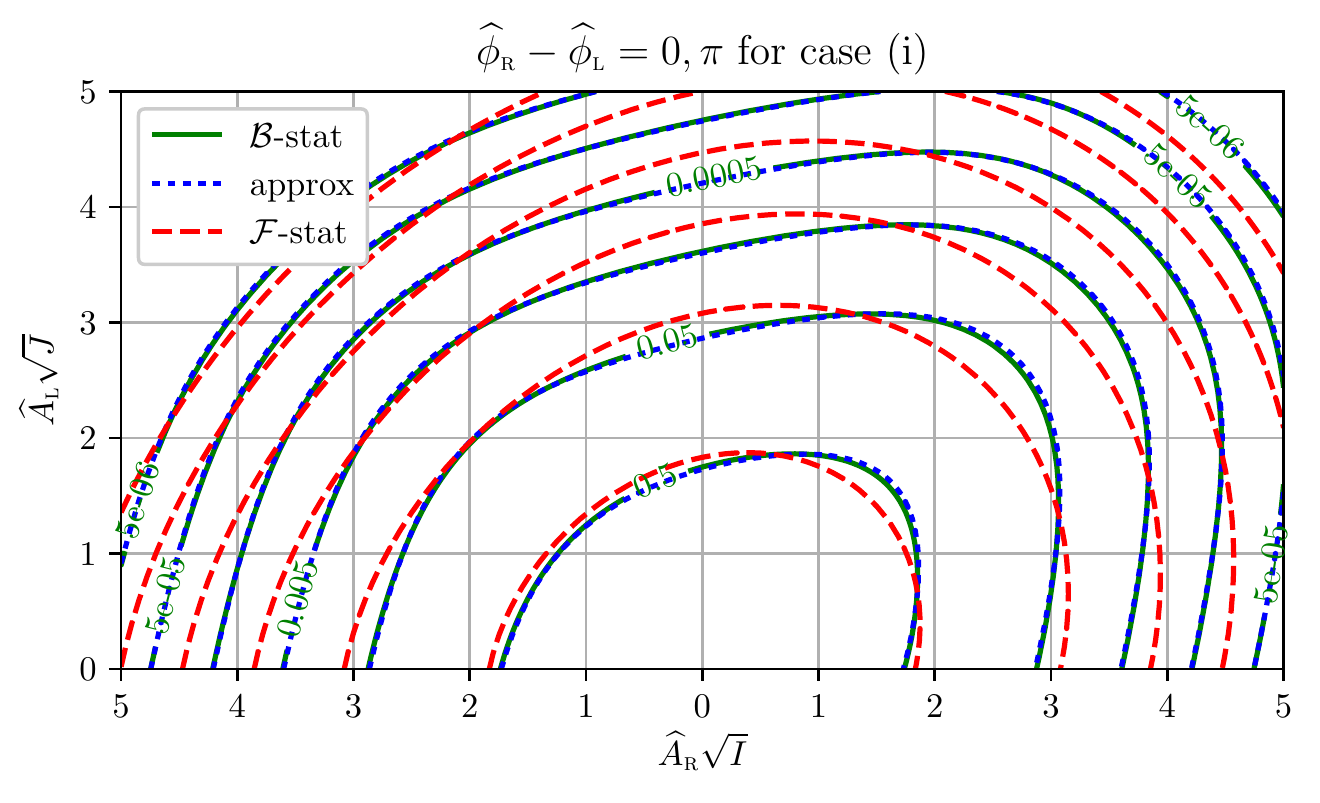}
\includegraphics[width=\textwidth]{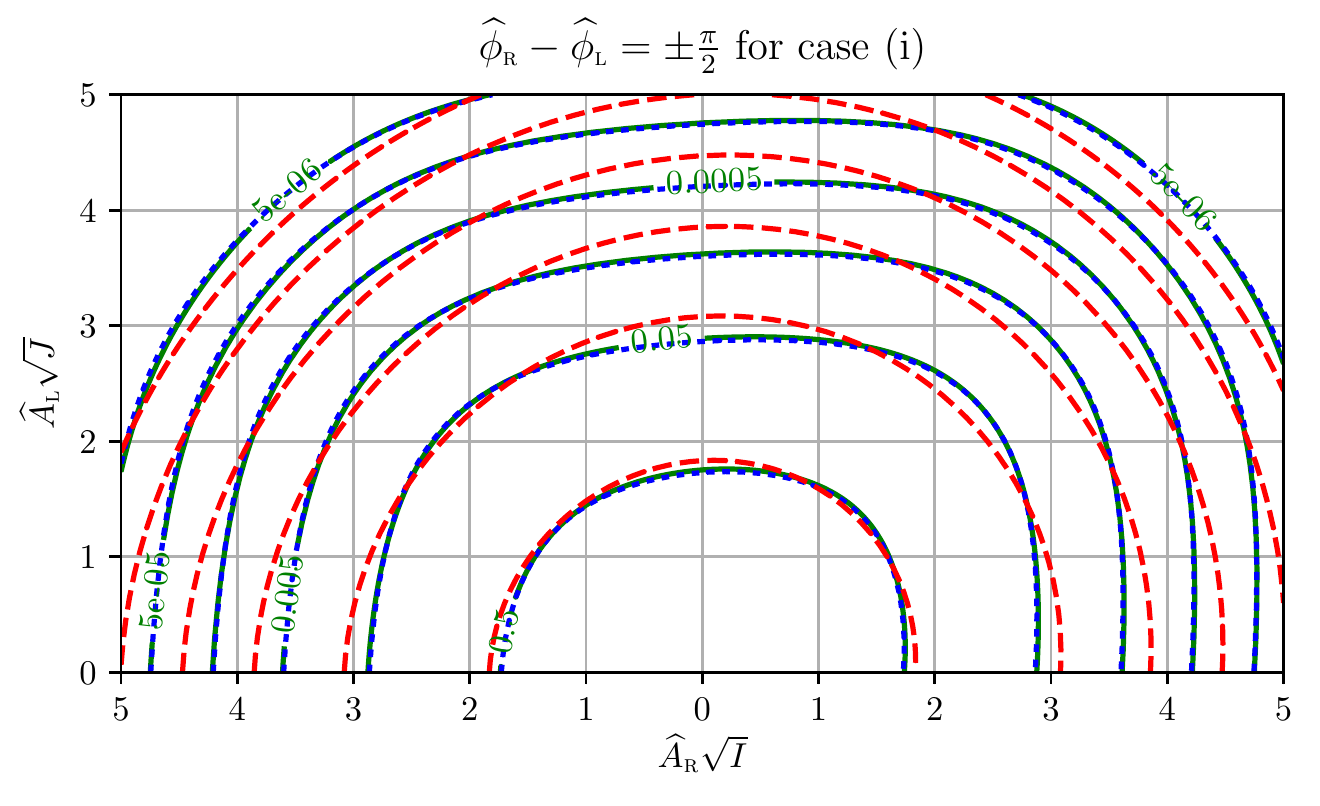}
\caption{Comparison of $\B$-statistic \eqref{e:Bstatexplicit} and
  approximation \eqref{e:BstatApprox}, along with $\F$-statistic
  \eqref{e:Fstat}; using the assumptions of a 25-hour observation
  beginning 2004 Jan 1 at 00:00 UTC (GPS time 756950413) [case (i)],
  for which $K/I=\PKKI$ and $L/I=\PKLI$.  The statistics depend on the
  data through the maximum-likelihood parameters $\MLEAR$, $\MLEAL$,
  and $\MLEtR-\MLEtL$.  Top: the slice $\sin(\MLEtR-\MLEtL)=0$, for
  which the $L$-dependent terms of the statistics are important;
  bottom: the slice $\cos(\MLEtR-\MLEtL)=0$, for which the
  $K$-dependent terms of the statistics are important.  The contours
  of constant exact and approximate statistic are nearly
  indistinguishable, indicating that this is a good approximation for
  these metric values.}
\label{f:PK_comparison_deg}
\end{center}
\end{figure}

We choose our contours for these plots to correspond to specific
false-alarm probabilities (estimated by drawing $10^7$ random points
$\{\MLE{\A}^{\mudot}\}$ from a Gaussian with zero mean and
variance-covariance matrix $\{\M^{\mudot\nudot}\}$) rather than
specific statistic values.  In \figref{f:PK_comparison_deg}, we see
that for the case (i), with $K/I=\PKKI$ and $L/I=\PKLI$, the
approximation works well and the approximate and exact $\B$-statistic
contours are nearly indistinguishable.
\begin{figure}
\begin{center}
\includegraphics[width=\textwidth]{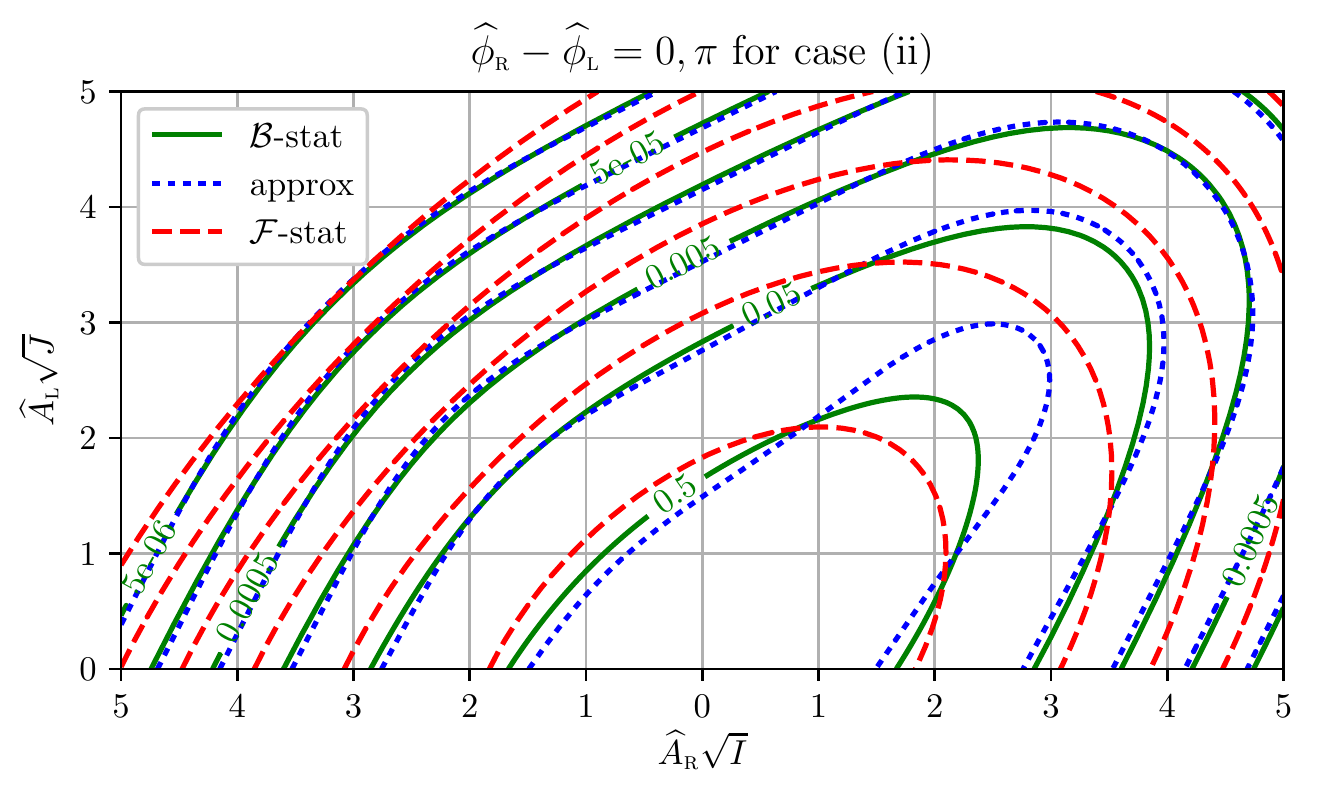}
\includegraphics[width=\textwidth]{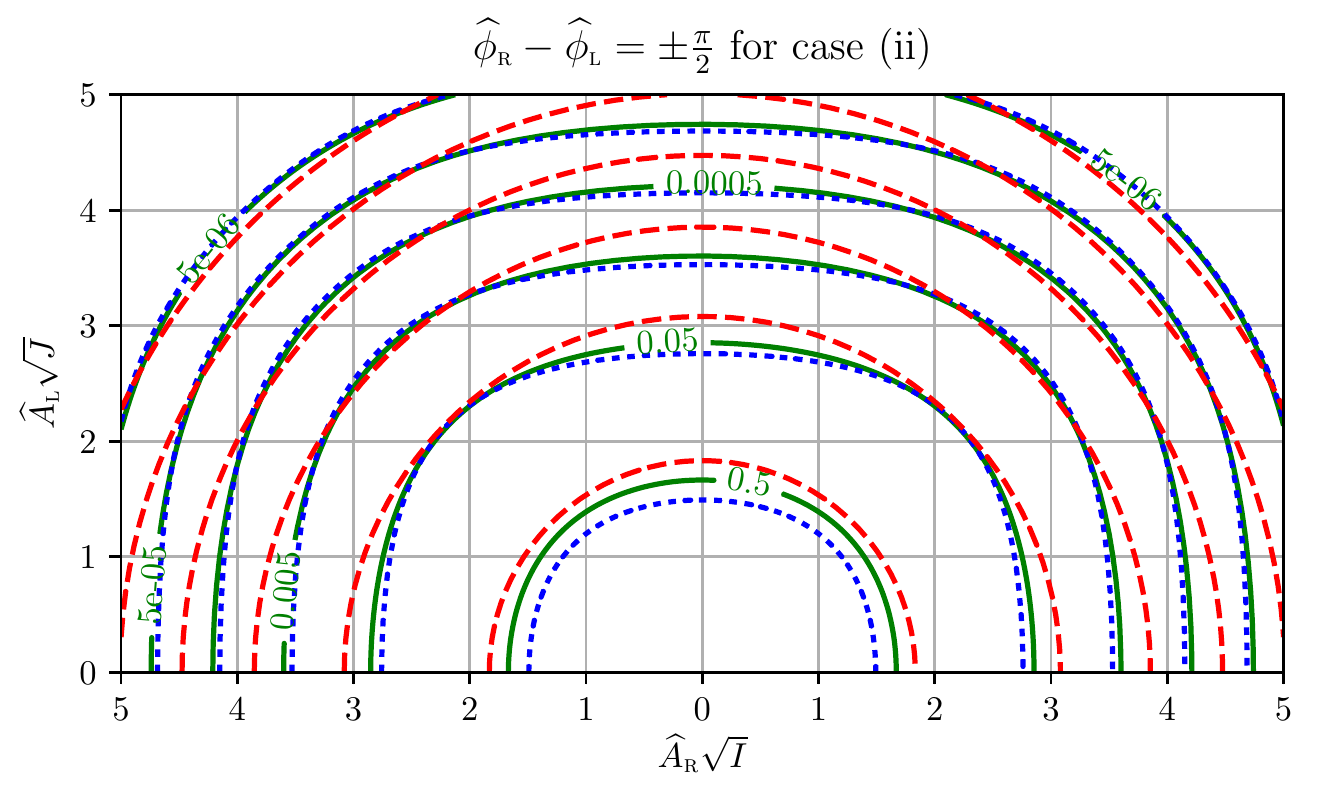}
\caption{Comparison of $\B$-statistic and approximation, along with
  $\F$-statistic, assuming a source on the celestial equator and H1
  observations which evenly sample sidereal time [case (ii)], for which
  $K/I=\HiKI$ and $L/I=\HiLI$, contours and slices constructed as in
  \protect\figref{f:PK_comparison_deg}.  There is some discrepancy
  between the approximate and exact $\B$-statistic contours at low
  false alarm rate in the case of linear polarization
  $\MLEAR\approx\MLEAL$.  Note that the disagreement for this contour
  in other directions is because it is drawn at the same false alarm
  probability, so the approximate $\B$-statistic contour must be
  inside the exact $\B$-statistic contour to compensate for the
  deformation in one direction.}
\label{f:H1_comparison_deg}
\end{center}
\end{figure}
\begin{figure}
\begin{center}
\includegraphics[width=\textwidth]{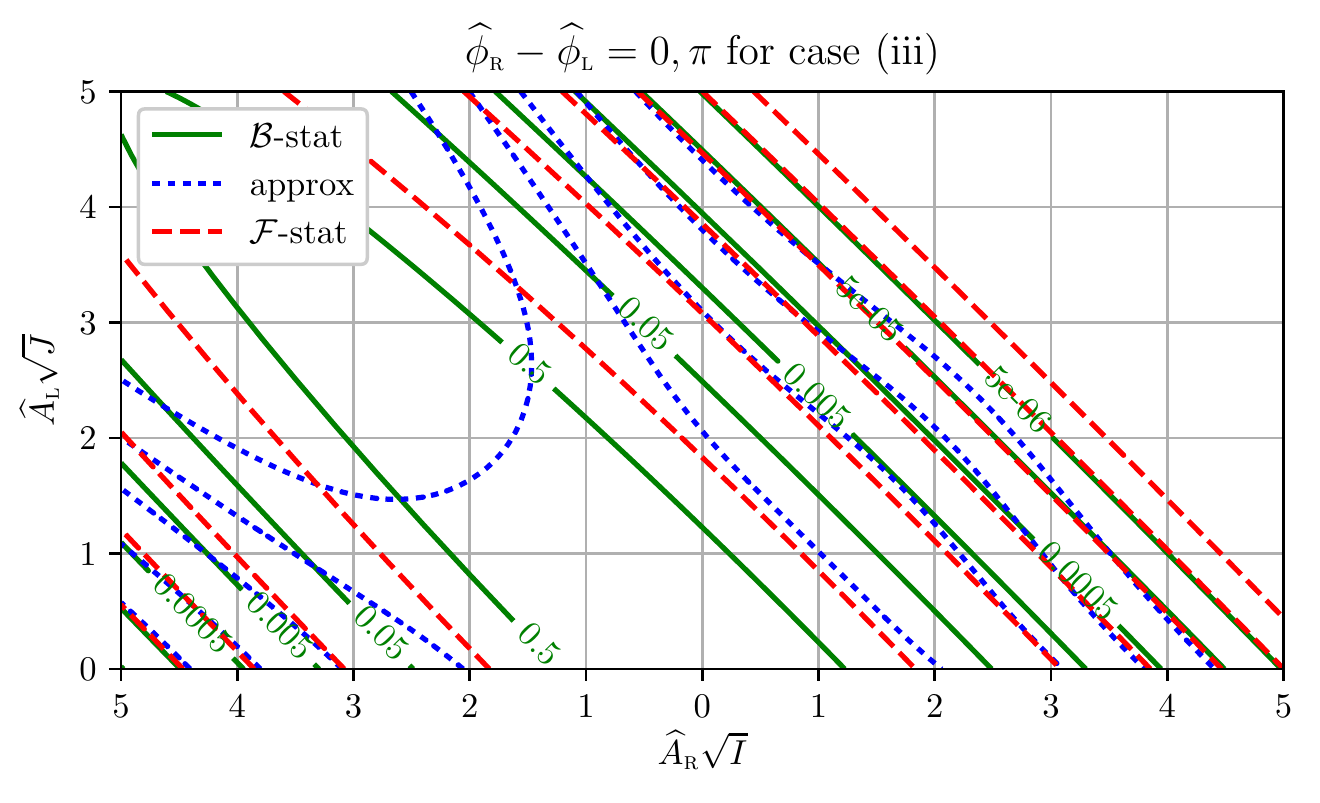}
\includegraphics[width=\textwidth]{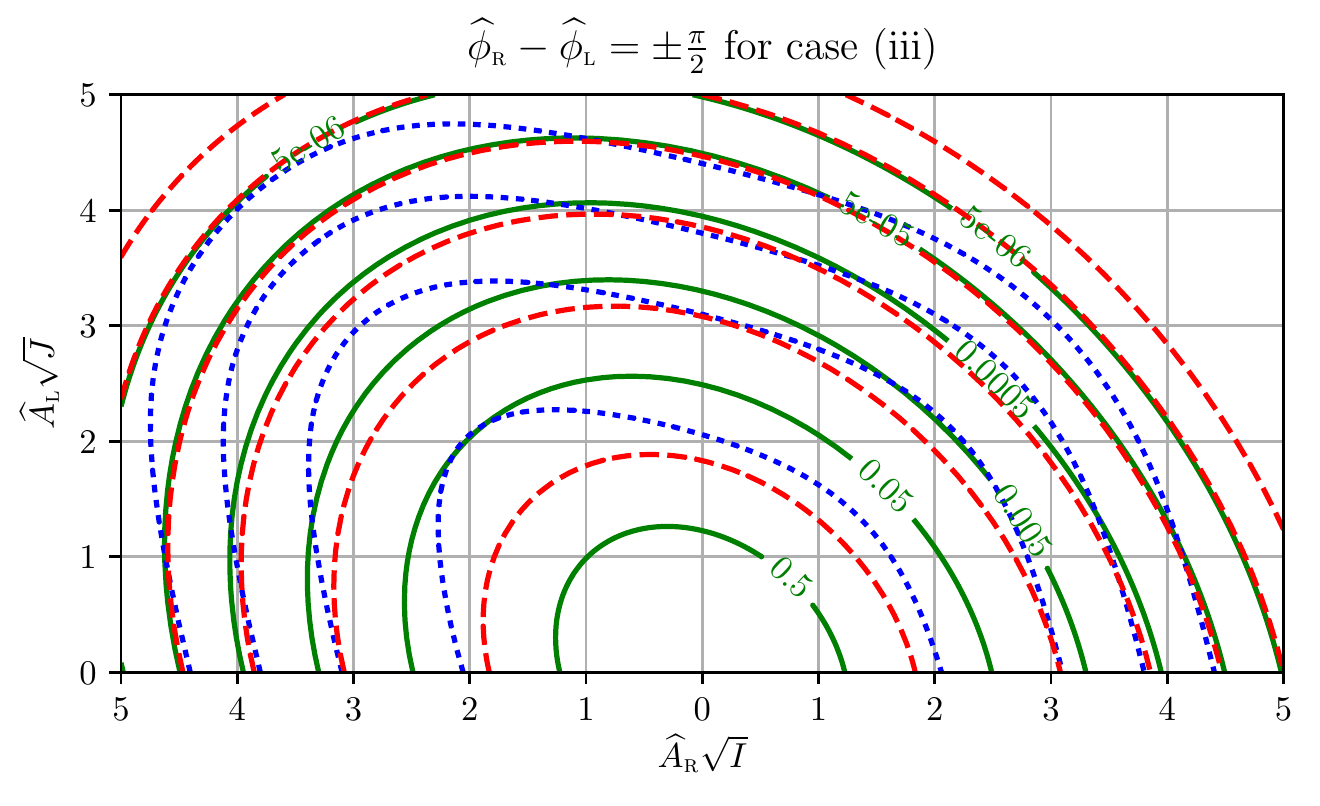}
\caption{Comparison of $\B$-statistic and approximation, along with
  $\F$-statistic, assuming a single Greenwich sidereal time of 00:00
  [case (iii)], for which $K/I=\oKI$ and $L/I=\oLI$, contours and
  slices constructed as in \protect\figref{f:PK_comparison_deg}.  Now
  the contours for the approximate $\B$-statistic are quite far off of
  those of the exact $\B$-statistic.  In fact, the entirety of both
  plots lie above the median of the approximate $\B$-statistic under
  the no-signal hypothesis; the contour in the upper left of the top
  plot is a false alarm probability of $.05$, and the one in the
  center of the lower plot is $.0005$.  The origin $\MLEAR=0=\MLEAL$
  is at the 98th percentile of the approximate $\B$-statistic, but the
  minimum of the exact $\B$-statistic.  Thus the approximation is, as
  expected, inappropriate for a value of $\sqrt{K^2+L^2}/I$ so close
  to unity.}
\label{f:GST0_comparison_deg}
\end{center}
\end{figure}
\Figref{f:H1_comparison_deg} shows case (ii), for which $K/I=\HiKI$
and $L/I=\HiLI$.  Some discrepancy is visible for low false-alarm
rates when the maximum-likelihood value corresponds to linear
polarization with $\MLE{\psi}\approx 0$, i.e.,
$\MLEAR e^{i\MLEtR}\approx\MLEAL e^{i\MLEtL}$.
Finally, in \figref{f:GST0_comparison_deg} we show the case (iii),
with $K/I=\oKI$ and $L/I=\oLI$.  The approximation performs badly, as
we'd expect for a first-order expansion in a quantity close to unity.

\begin{figure}
\begin{center}
\includegraphics[width=\textwidth]{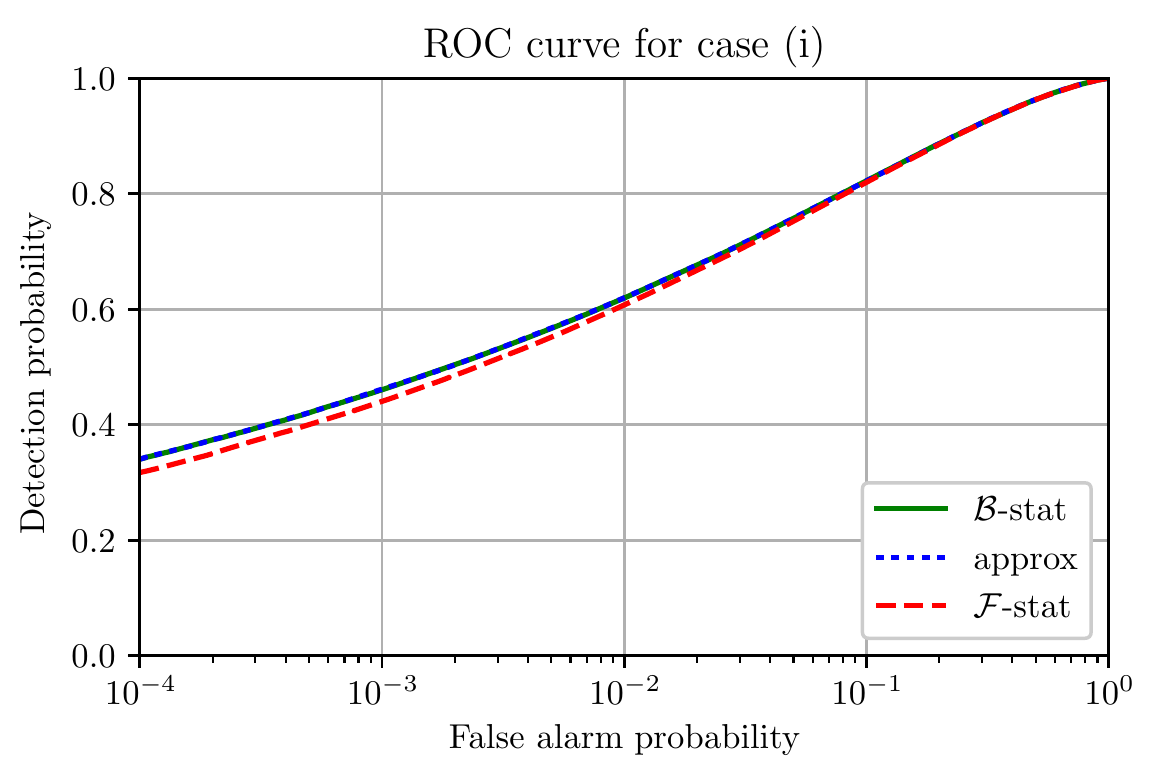}
\caption{ROC curves for $\B$-statistic and approximation, along with
  $\F$-statistic, using the metric from case (i) (see
  \protect\figref{f:PK_comparison_deg}).  In this case, the
  approximate $\B$-statistic performs identically to the exact one.
  Compare figure~3 of \cite{PK}.}
\label{f:PK_ROC}
\end{center}
\end{figure}

\subsection{Monte Carlo Simulations}

To evaluate the performance of our $\B$-statistic approximation, we
produced Monte Carlo simulations by drawing $10^6$ sets of signal
parameters, using a fixed value of $h_0=10\frac{S_n(f_0)}{\Tobs}$ and
drawing the parameters $\cosi$, $\psi$, and $\phi_0$ from uniform
distributions.  Each of these sets of parameters was converted into a
point $\A^{\mudot}$, and then a signal $\MLE{\A}^{\mudot}$ was
generated by drawing from a Gaussian with mean $\A^{\mudot}$ and
variance-covariance matrix $\{\M^{\mudot\nudot}\}$.  A
receiver-operating-characteristic (ROC) curve was generated for each
statistic by plotting the fraction of signal points above a signal
threshold (detection probability) against the fraction of noise points
(described in the previous section) above the same threshold.  The
latter fraction is known as false-alarm probability, Type I error
probability, or, in the language of hypothesis testing, significance.
A superior detection statistic will have a higher detection efficiency
at a given false-alarm probability, and thus be found above and to the
left of an inferior one.  Note that while the Neyman-Pearson lemma
states that the Bayes factor will be the optimal test statistic for a
Monte Carlo using the same prior\cite{2008arXiv0804.1161S} this is not
guaranteed to be the case here, since the delta-function prior on
$h_0$ is not the same as the uniform prior used in defining the
statistic.

\begin{figure}
\begin{center}
\includegraphics[width=\textwidth]{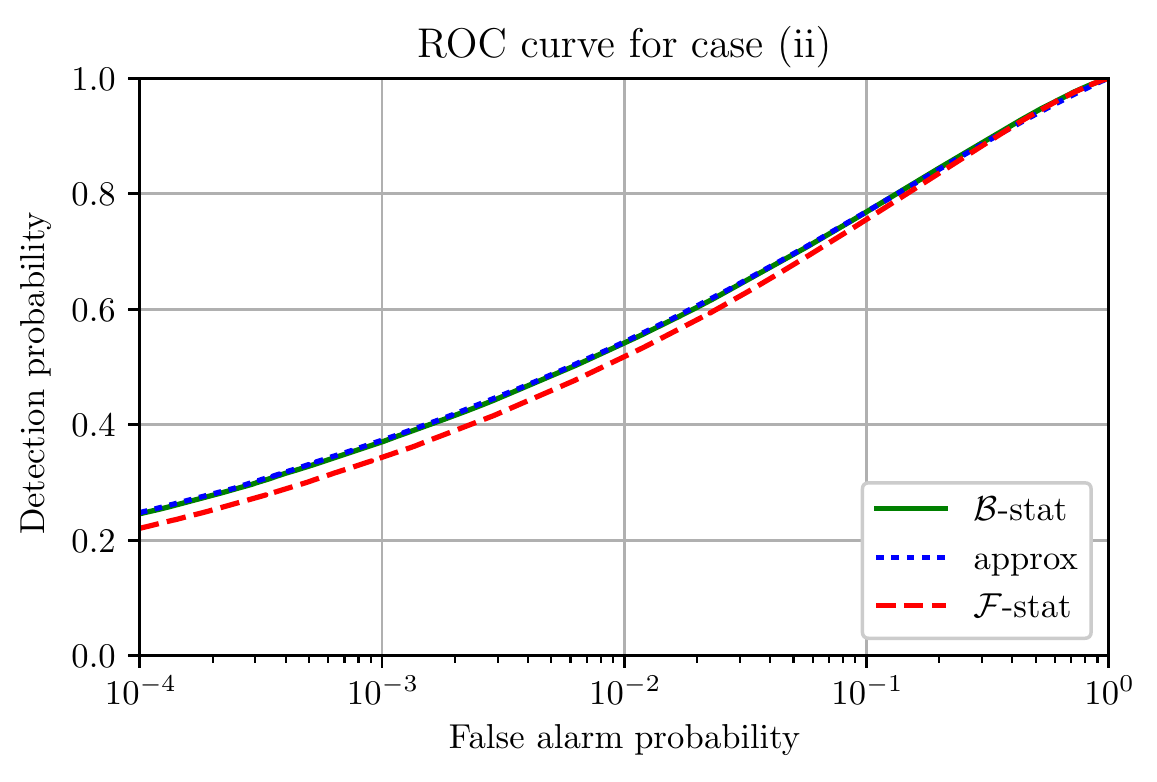}
\caption{ROC curves for $\B$-statistic and approximation, along with
  $\F$-statistic, using the metric from case (ii) (see
  \protect\figref{f:H1_comparison_deg}).  Even though $K/I=\HiKI$ and
  $L/I=\HiLI$, the approximate $\B$-statistic, which is Taylor expanded in
  $K/I$ and $L/I$, still performs as well as the exact  $\B$-statistic (and
  better than the $\F$-statistic) in this Monte Carlo.}
\label{f:H1_ROC}
\end{center}
\end{figure}

\begin{figure}
\begin{center}
\includegraphics[width=\textwidth]{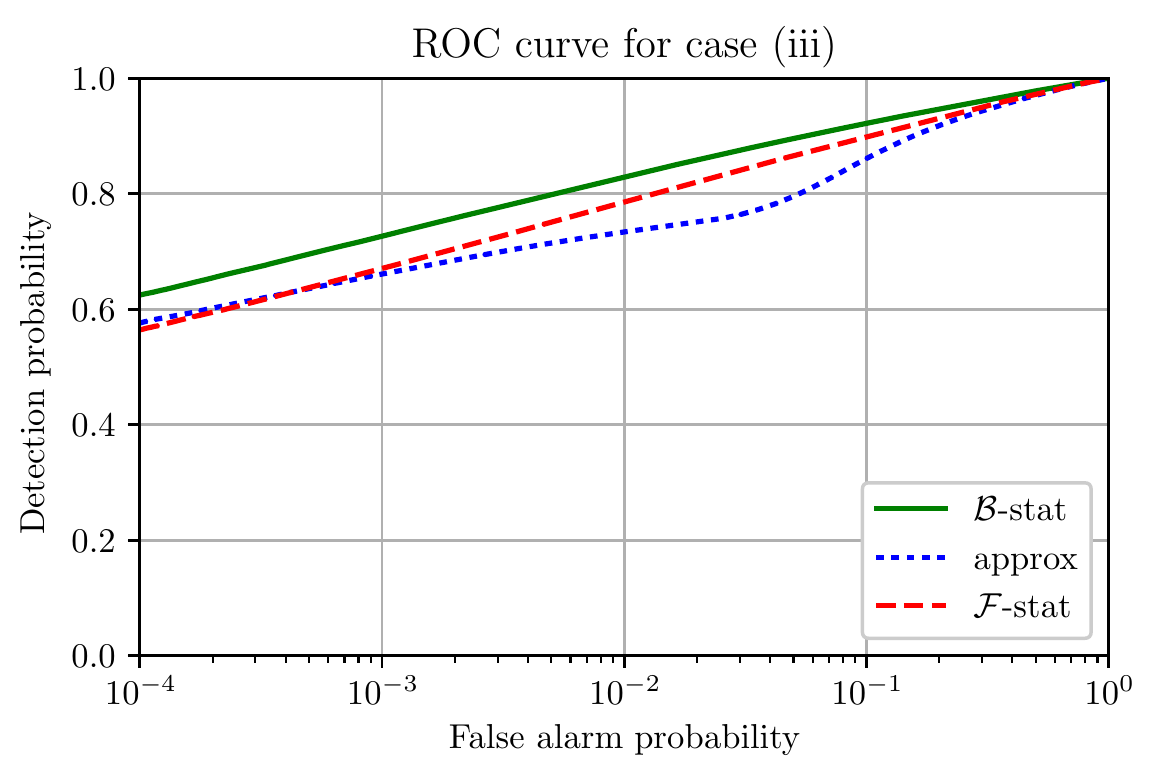}
\caption{Comparison of $\B$-statistic and approximation, along with
  $\F$-statistic, using the metric from case (iii) (see
  \protect\figref{f:GST0_comparison_deg}).  Here $K/I=\oKI$ and
  $L/I=\oLI$, and we see indeed that the approximate $\B$-statistic
  performs poorly, considerably below both the exact $\B$-statistic
  and the $\F$-statistic.}
\label{f:GST0_ROC}
\end{center}
\end{figure}

In \figref{f:PK_ROC} we show the ROC curve for case (i), in which our
approximation was shown to match the exact $\B$-statistic well (see
\figref{f:PK_comparison_deg}).  As expected, the approximate
$\B$-statistic performs as well as the exact one, and both outperform
the $\F$-statistic, as shown in \cite{PK}.
In \figref{f:H1_ROC} we show the ROC curve for case (ii), where our
approximation was shown in \figref{f:H1_comparison_deg} to have some
discrepancies with the exact $\B$-statistic.  Nonetheless, we see that
it again performs as well as the exact $\B$-statistic and better than
the $\F$-statistic.
In \figref{f:GST0_ROC} we show the ROC curve for case (iii), where our
approximation was shown in \figref{f:GST0_comparison_deg} to disagree
considerably with the exact $\B$-statistic.  Unsurprisingly, we find
this approximation to be a poor detection statistic in this scenario,
underperforming both the exact $\B$-statistic and the $\F$-statistic.

\section{Conclusions}
\label{s:conclusions}

We have produced an analytic approximation to the $\B$-statistic, a
Bayesian detection statistic for continuous gravitational waves based
on a Bayes factor between signal and noise hypotheses.  This
approximation is based on a Taylor expansion in the parameters $K/I$
and $L/I$, which are related to observation-averaged combinations of
antenna patterns, and depend on the sky position of the source,
detectors involved in the observation, and distribution of the
observations in sidereal time.  For long-time observations which
average over a range of sidereal times, these parameters tend to be
small enough to produce a good first-order approximation, and we
showed via Monte Carlo simulations that the approximate statistic
performed as well as the exact $\B$-statistic, even for a case with an
expansion parameter approaching 50\%.  The approximation is shown to
break down for observations at a single sidereal time, which indicates
the approximation is not likely to be an appropriate statistic for
transient modelled signals such as compact binary inspiral.

Unlike the exact $\B$-statistic, which must be evaluated via a
two-dimensional numerical integral, our approximation (like the
maximum-likelihood $\F$-statistic) can be evaluated analytically,
which should make it computationally more efficient.\footnote{For
  example, for the python code used to perform the Monte Carlo
  simulations for this paper, the signel code used to calculate both the
  $\F$-statistic and the approximate $\B$-statistic together took
  $\mc{O}(1-2\,\un{\textmu{s}})$ per evaluation, while the numerical
  integration for the exact $\B$-statistic took $\mc{O}(1\un{ms})$ per
  evaluation.}
This, combined with the better detection efficiency than the
$\F$-statistic at the same false alarm rate, makes it a potentially
useful replacement for, or alternative to, the $\F$-statistic in a
semicoherent search which combines $\F$-statistic values at a range of
signal parameters.  One potential challenge is that the approximate
$\B$ statistic is expressed in terms of confluent hypergeometric
functions, which may be more time-consuming to evaluate than the
algebraic functions involved in the $\F$-statistic.  Additionally,
direct evaluation of these confluent hypergeometric functions for
large-amplitude signals can produce overflow, even though the final
approximation in terms of their logarithms and ratios may be
well-behaved.  It may be necessary to supplement standard library
functions with strategic use of asymptotic forms.

\section*{Acknowledgments}

We wish to thank the members of the LIGO Scientific Collaboration and
Virgo Collaboration continuous waves group for useful feedback.
This paper has been assigned LIGO Document Number \dcc.

\appendix

\section{Form and Behavior of the Metric Elements}

\label{app:metric}

Given some nearly monochromatic GW signal around frequency $f_0$, the
multi-detector scalar product of two time series $x$ and $y$, used in
the definition \eqref{eq:Mmunu}, can be expressed as
\begin{equation}
  \scalar{x}{y} \equiv \sumXiSFT \frac{4}{S^X_\iSFT(f_0)}
  \Real\int_{0}^{\infty} \cft{x}^{X*}_\iSFT(f) \, \cft{y}^{X}_\iSFT(f) \,df
\ ,
  \label{eq:def-scalar}
\end{equation}
where $S^X_\iSFT(f_0)$ is the one-sided noise power spectral density
around the frequency $f_0$ in detector $X$ during time stretch
$\iSFT$, and $\cft{x}_\iSFT^X(f), \cft{y}_\iSFT^X(f)$ are the
corresponding Fourier-transforms of $x^X(t),y^X(t)$ restricted to the
time stretch $\iSFT$. This assumes the data from each detector $X$ has
been divided into short stretches of data
$[t_\iSFT,\,t_\iSFT + \Tsft)$ of length $\Tsft$.

The metric components can be written as
\begin{equation}
  I = A + B + 2E \quad\hbox{and}\quad
  J = A + B - 2E \quad\hbox{and}\quad
  K = 2C \quad\hbox{and}\quad
  L = A - B
\end{equation}
where, in the long-wavelength limit,
\begin{equation}
  \label{e:ABC_LWL}
  A = \sumXiSFT\frac{\Tsft}{S^X_\iSFT(f_0)}\left(a^X_\iSFT\right)^2
  \quad\hbox{and}\quad
  B = \sumXiSFT\frac{\Tsft}{S^X_\iSFT(f_0)}\left(b^X_\iSFT\right)^2
  \quad\hbox{and}\quad
  C = \sumXiSFT\frac{\Tsft}{S^X_\iSFT(f_0)}a^X_\iSFT b^X_\iSFT
\end{equation}
and $E=0$ (so that $I=J$).  As shown in \cite{WPCW}, the more general
expression, with a complex frequency-dependent detector tensor
$\tens{d}(f)$ and amplitude-modulation {\coeff}s $a(f)$ and $b(f)$,
the (real) metric components can be more generally written
as\footnote{Note that equation (A.3b) of \cite{WPCW} has the formulas
  for $K$ and $L$ reversed.}
\begin{subequations}
  \begin{gather}
    I = \sumXiSFT\frac{\Tsft}{S^X_\iSFT(f_0)}
    \left\lvert a^X_\iSFT(f_0)-ib^X_\iSFT(f_0)\right\rvert^2
    \quad\hbox{and}\quad
    J = \sumXiSFT\frac{\Tsft}{S^X_\iSFT(f_0)}
    \left\lvert a^X_\iSFT(f_0)+ib^X_\iSFT(f_0)\right\rvert^2
    \\
    L + iK = \sumXiSFT\frac{\Tsft}{S^X_\iSFT(f_0)}
    \left[a^X_\iSFT(f_0)-ib^X_\iSFT(f_0)\right]^*
    \left[a^X_\iSFT(f_0)+ib^X_\iSFT(f_0)\right]
  \end{gather}
\end{subequations}
In this form, we can see that the Cauchy-Schwarz inequality implies
that
\begin{equation}
  K^2 + L^2 = \abs{L+iK}^2 \le IJ
  \ ;
\end{equation}
in the long-wavelength case, this becomes $\sqrt{K^2+L^2}\le I=J$.

\begin{figure}
\begin{center}
\includegraphics[width=0.48\textwidth]{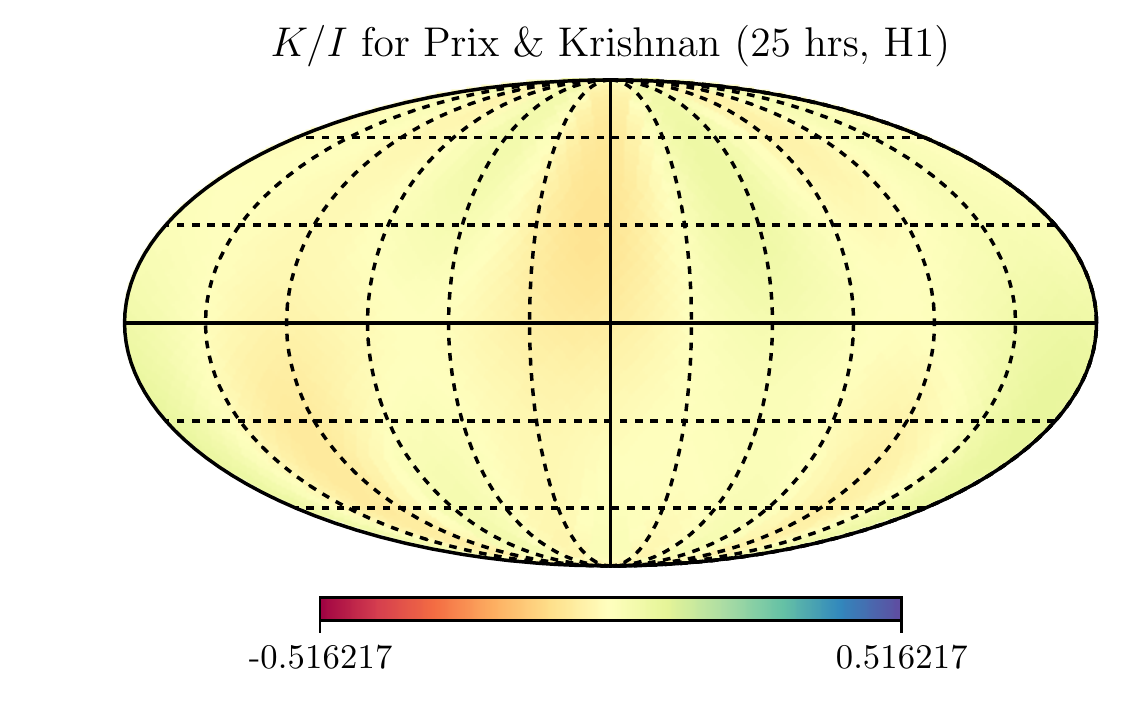}
\includegraphics[width=0.48\textwidth]{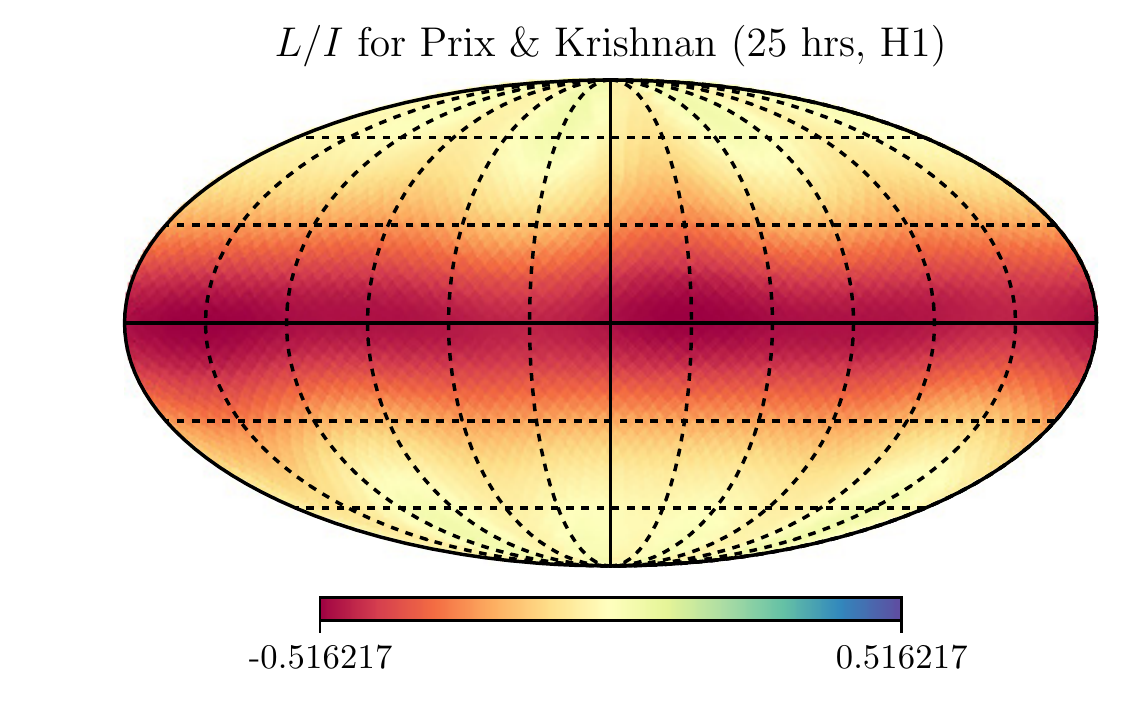}
\includegraphics[width=0.48\textwidth]{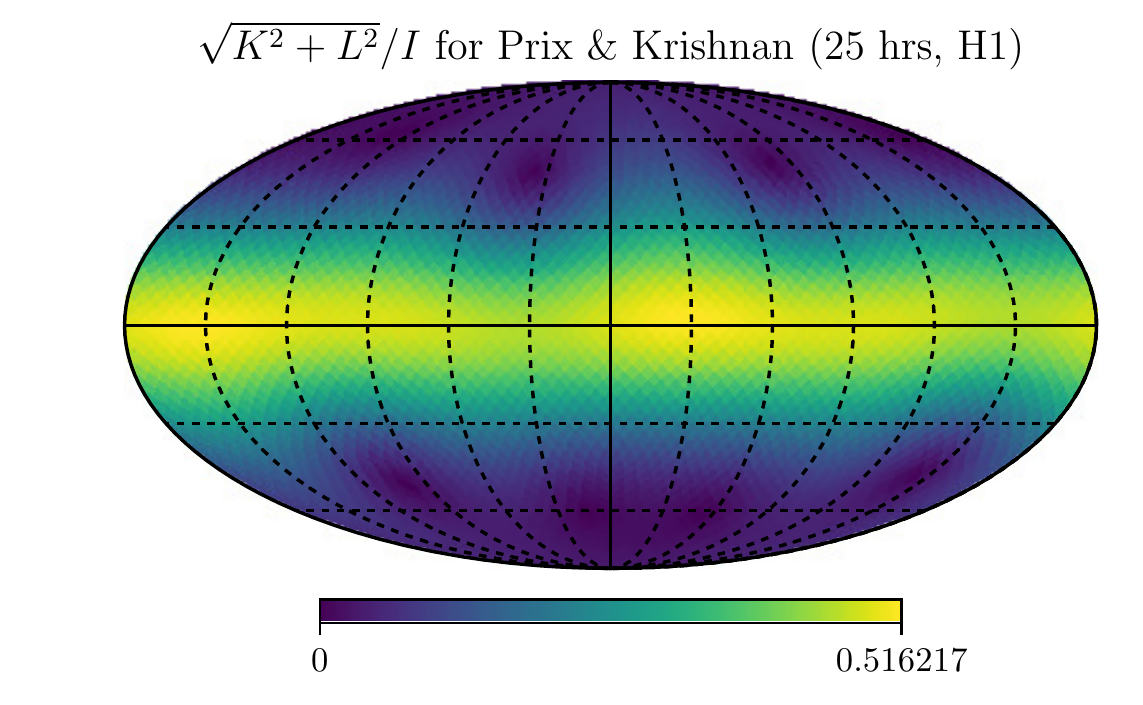}
\includegraphics[width=0.48\textwidth]{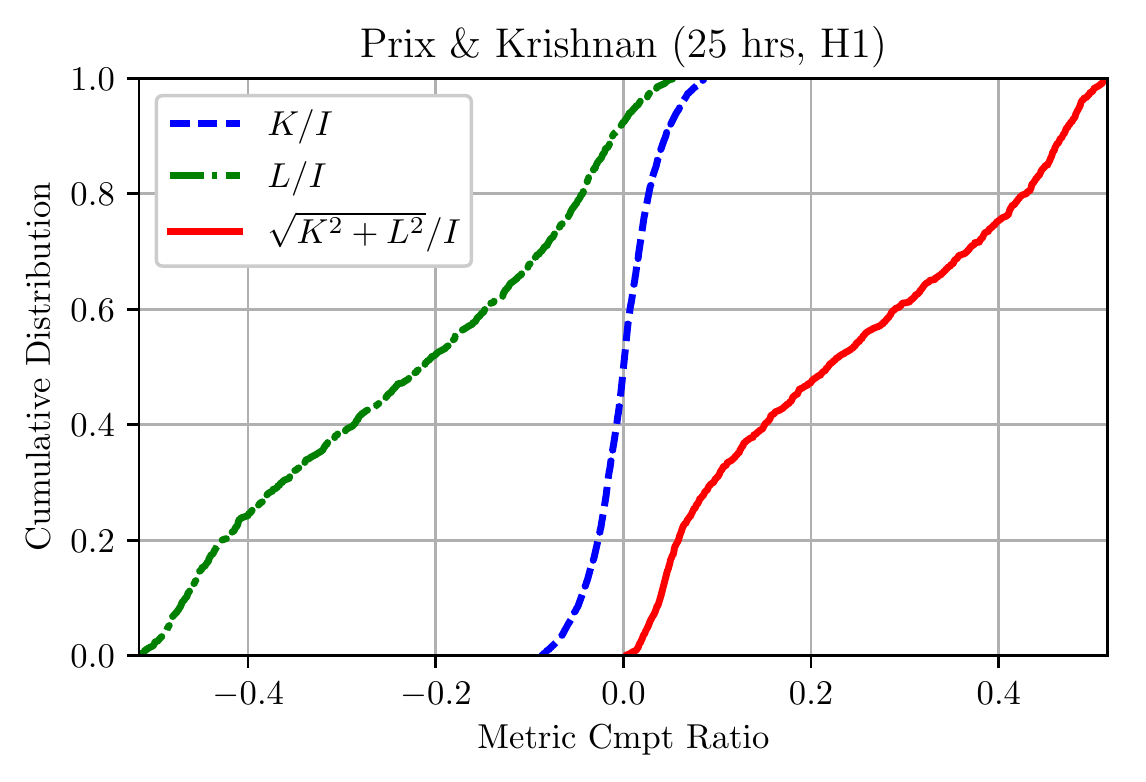}
\caption{Plots of the metric element ratios $K/I$, $L/I$, and
  $\sqrt{K^2+L^2}/I$ versus sky position of targeted source, along
  with cumulative probability distributions of these ratios, assuming a
  randomly chosen sky location, using the assumption of a 25-hour
  observation with LIGO Hanford Observatory (H1) beginning 2004 Jan 1
  at 00:00 UTC (GPS time 756950413).}
\label{f:PK_sky}
\end{center}
\end{figure}

Prix and Krishnan \cite{PK}
give an example of a $\Tobs=25\un{hr}$ observation of a source at
right ascension $2\un{radians}$, declination $-0.5\un{radians}$, with a
single detector (LIGO Hanford, known as H1) beginning at GPS time
756950413 (2014 Jan 1 at 00:00 UTC) and obtain metric components of
values of $A=\PKA\frac{\Tobs}{S_n(f_0)}$,
$B=\PKB\frac{\Tobs}{S_n(f_0)}$, and $C=\PKC\frac{\Tobs}{S_n(f_0)}$,
which is equivalent to $I=J=\PKI\frac{\Tobs}{S_n(f_0)}$,
$K=\PKK\frac{\Tobs}{S_n(f_0)}$, and $L=\PKL\frac{\Tobs}{S_n(f_0)}$. or
$K/I=\PKKI$, $L/I=\PKLI$.  We explore the robustness of those ratios
in \figref{f:PK_sky}, which calculated them for the same observing
time and different sky positions.  The ratio $K/I$ is small ($<0.10$)
everywhere, while the ratio $L/I$ is smaller away from the celestial
equator.

As an alternative to the arbitrarily chosen 25-hour observing time of
\cite{PK}, we can consider the idealization that a long
observation will include roughly the same amount of data from each
sidereal time, and construct the corresponding metric components for
this case.  Under this idealization, the metric components will be
independent of right ascension, allowing us to simply plot them versus
declination.  In \figref{f:skyavg} we plot the metric elements and
their ratios versus declination.
\begin{figure}
\begin{center}
\includegraphics[width=0.48\textwidth]{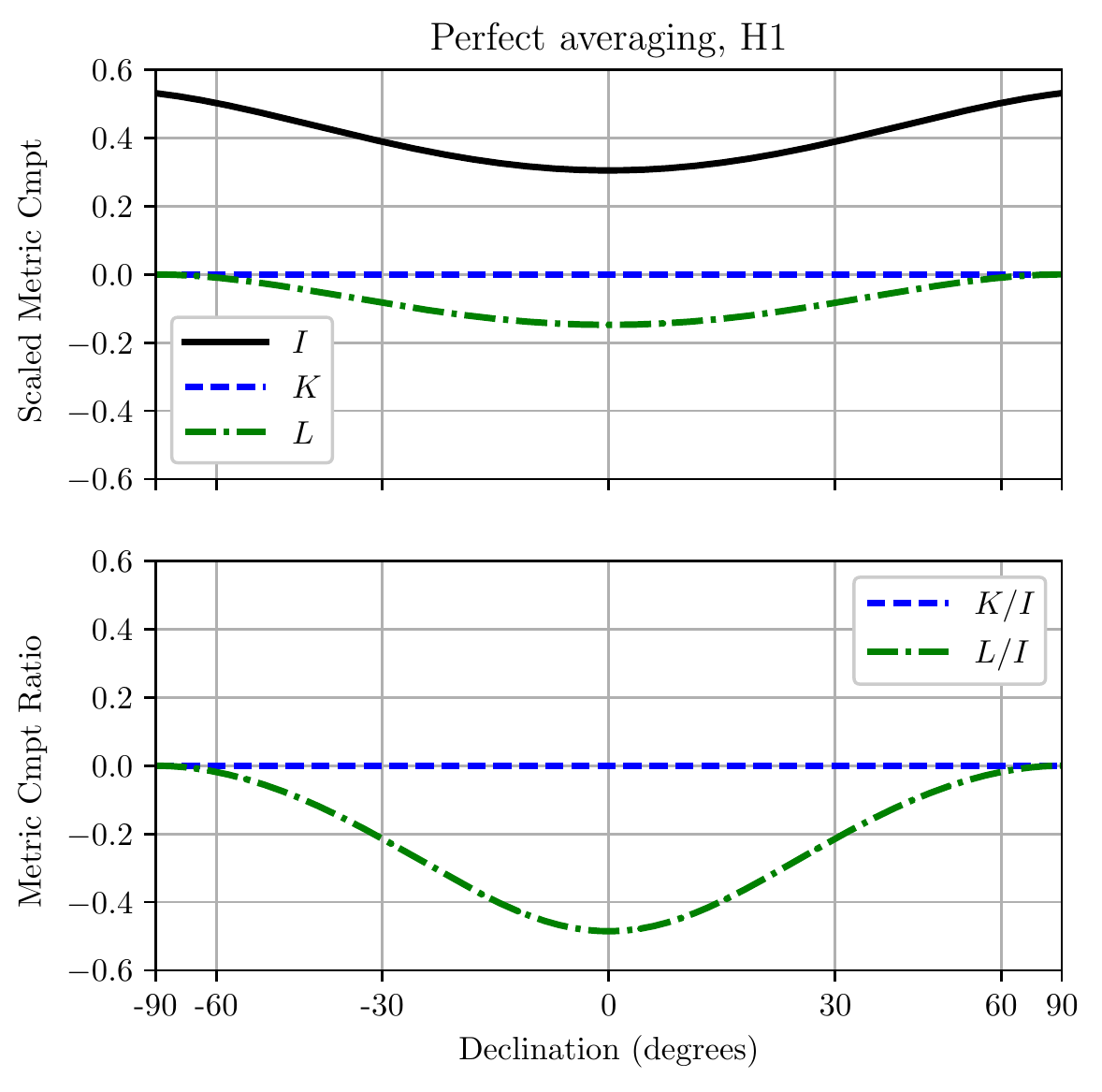}
\includegraphics[width=0.48\textwidth]{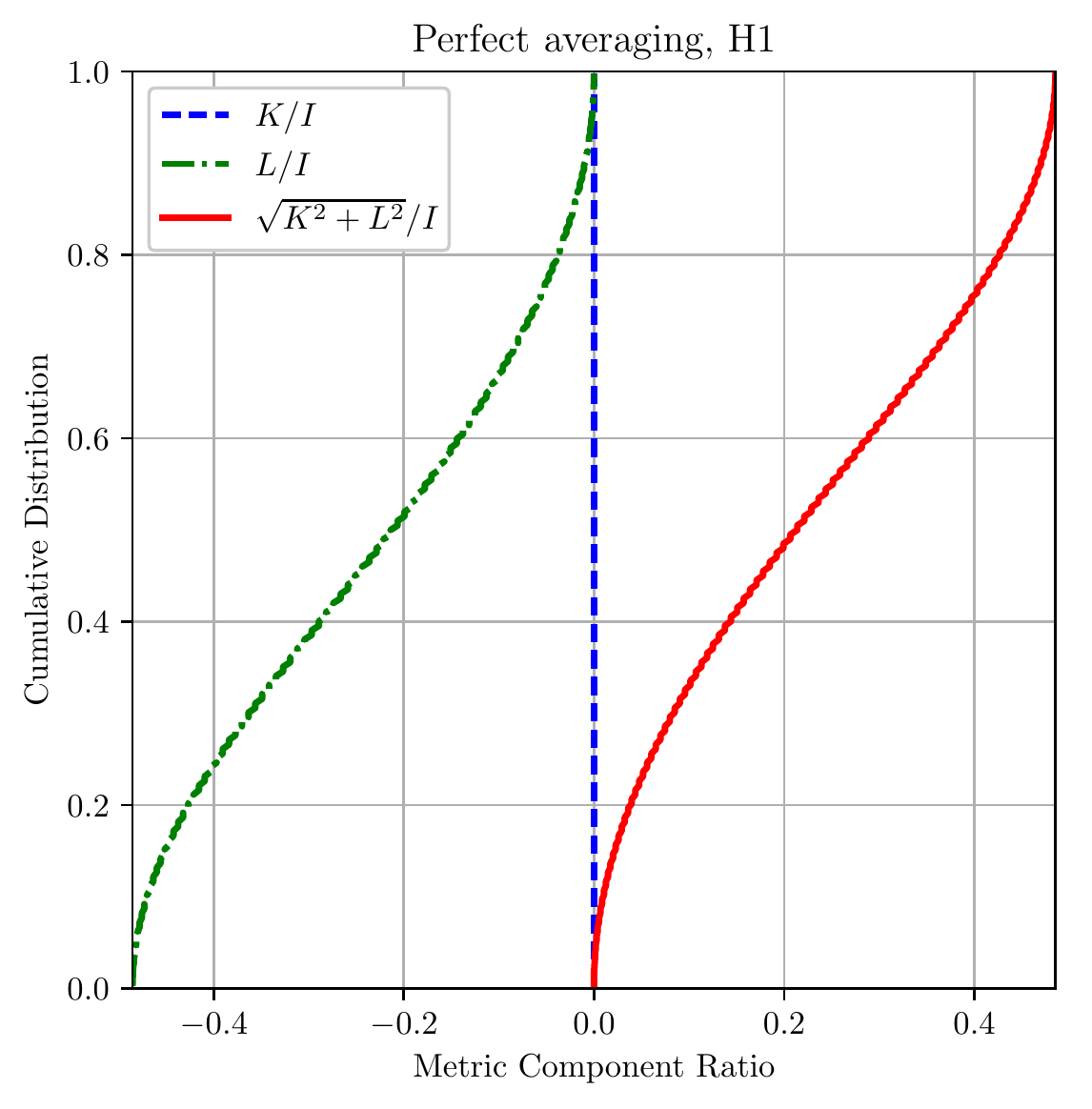}
\caption{Left: Plots of metric elements $I$, $K$, and $L$, and the
  ratios $K/I$ and $L/I$ versus declination of targeted source,
  assuming an observation using LIGO Hanford Observatory (H1) that
  results in a perfect average over sidereal time.  The spacing in
  declination is chosen to be proportional to sky area.  Right:
  Cumulative probability distributions of the metric element ratios
  $K/I$, $L/I$, and $\sqrt{K^2+L^2}/I$ for this case, assuming a
  randomly chosen sky location.}
\label{f:H1_skyavg}
\end{center}
\end{figure}
\begin{figure}
\begin{center}
\includegraphics[width=0.48\textwidth]{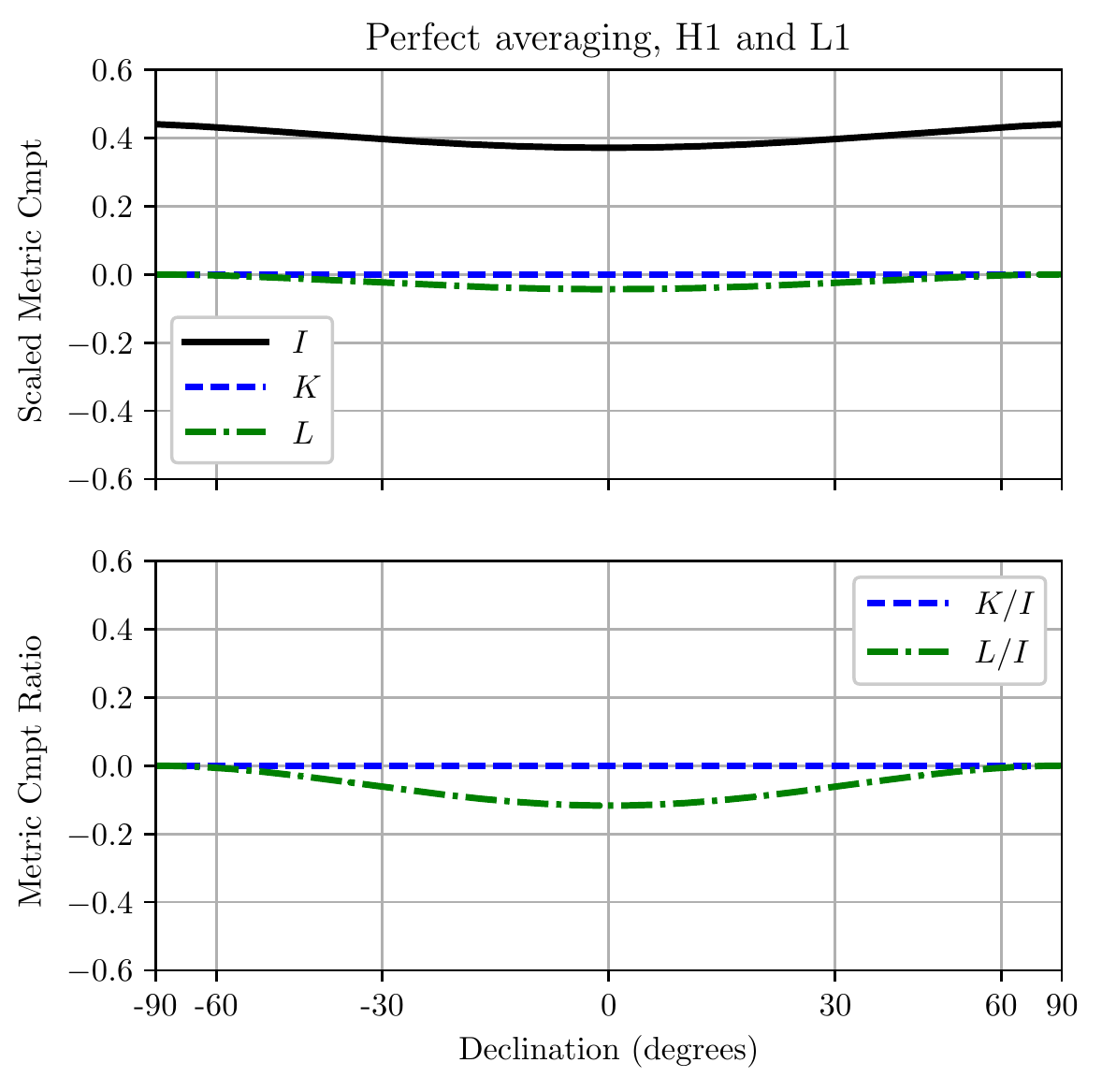}
\includegraphics[width=0.48\textwidth]{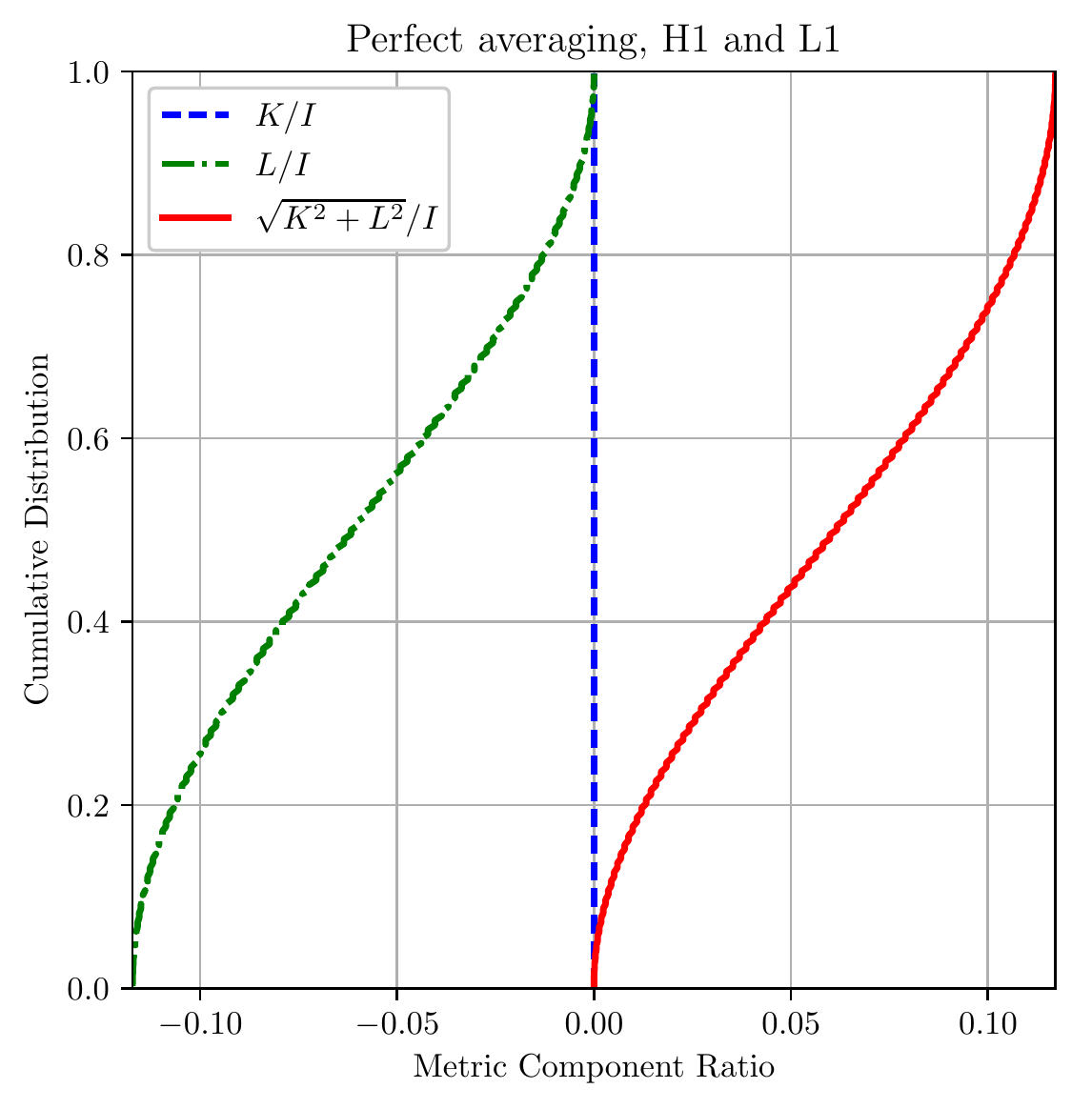}
\caption{Left: Plots of metric elements $I$, $K$, and $L$, and the
  ratios $K/I$ and $L/I$ versus declination of targeted source,
  assuming an observation using LIGO Hanford Observatory (H1) and LIGO
  Livingston Observatory (L1) that results in a perfect average over
  sidereal time.  The spacing in declination is chosen to be
  proportional to sky area.  Right: Cumulative probability
  distributions of the metric element ratios $K/I$, $L/I$, and
  $\sqrt{K^2+L^2}/I$ for this case, assuming a randomly chosen sky
  location.}
\label{f:skyavg}
\end{center}
\end{figure}
We find, as in \figref{f:PK_sky}, the ratio $L/I$ can approach $0.50$
near the celestial equator.  However, this is specific to the choice
of single-detector observations with H1 only.  If we assume equal
amounts of data from LIGO Hanford (H1) and LIGO Livingston (L1), we
find that $L/I\lesssim 0.15$ over the entire sky.  We also notice that
$K=0$ for this choice of observing time.  This is a geometrical result
related to the symmetries of the quantity $a^X b^X$ under rotations of
the Earth.

To give a more realistic example of a typical observing time, we
consider the H1 and L1 segments associated with advanced LIGO's first
observing run (O1)\footnote{https://doi.org/10.7935/K57P8W9D}, from
the LIGO Open Science Center\cite{Vallisneri_LOSC}.
\begin{figure}
\begin{center}
\includegraphics[width=0.48\textwidth]{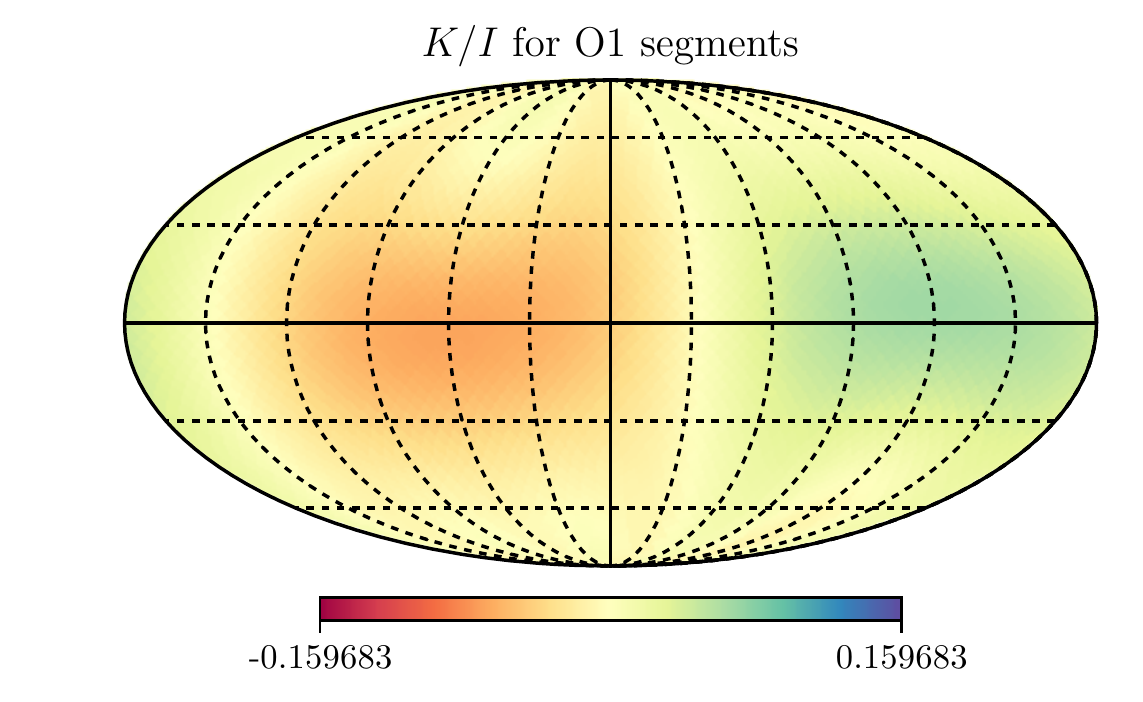}
\includegraphics[width=0.48\textwidth]{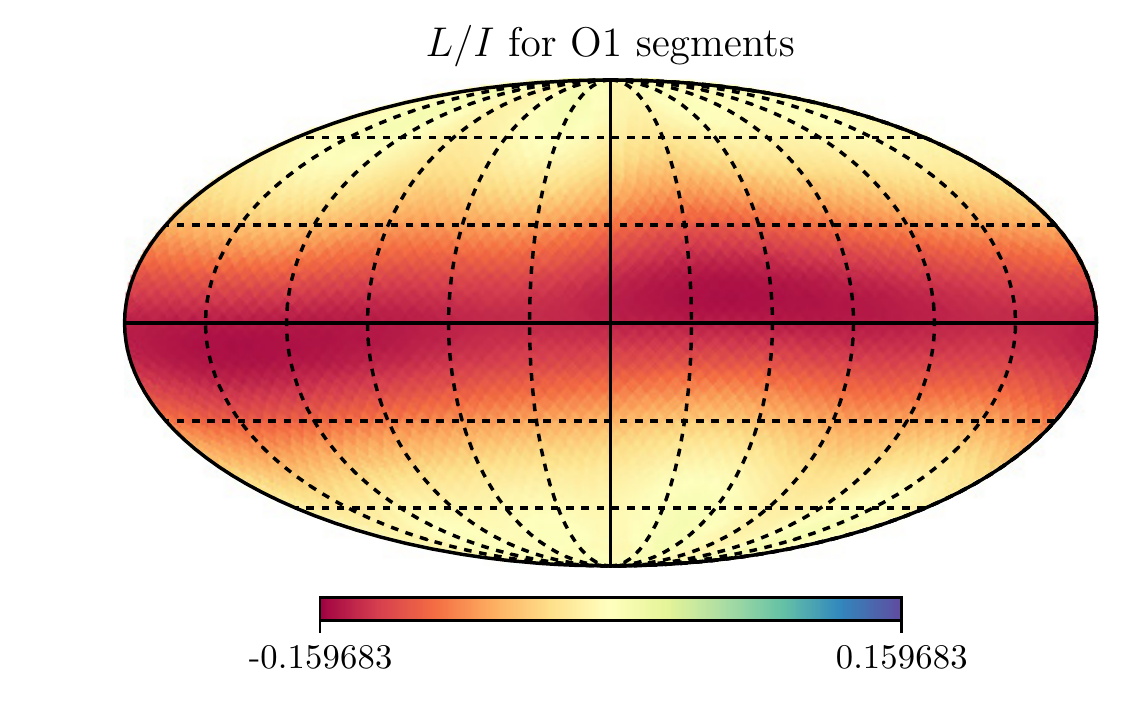}
\includegraphics[width=0.48\textwidth]{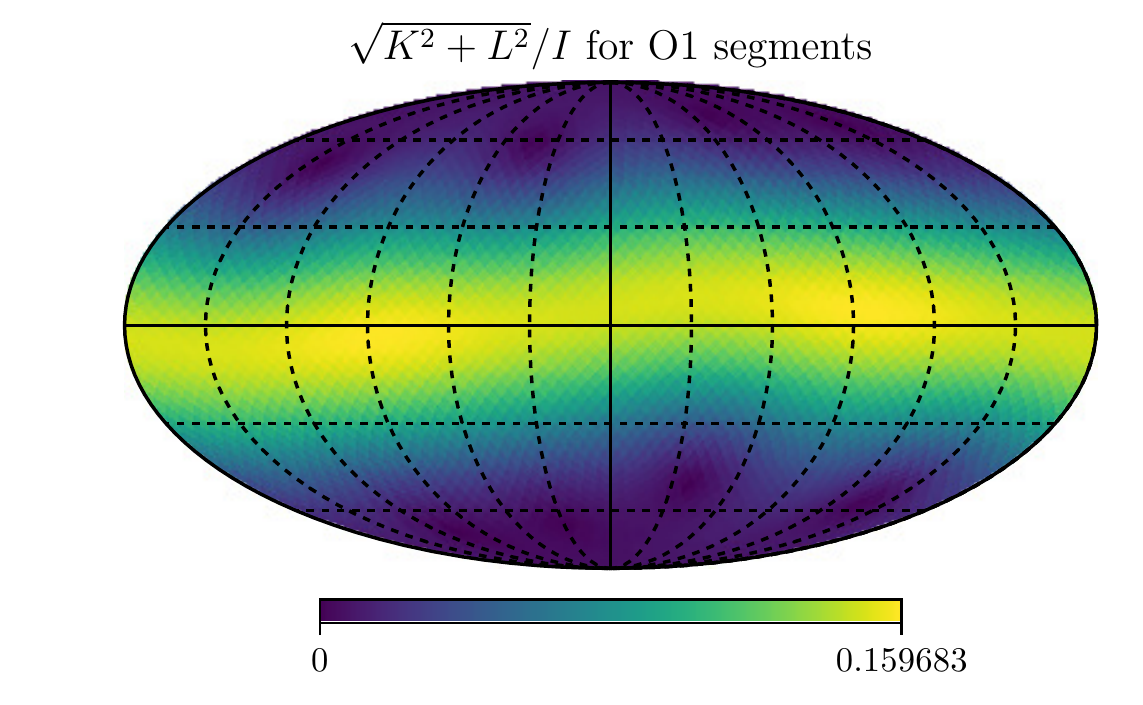}
\includegraphics[width=0.48\textwidth]{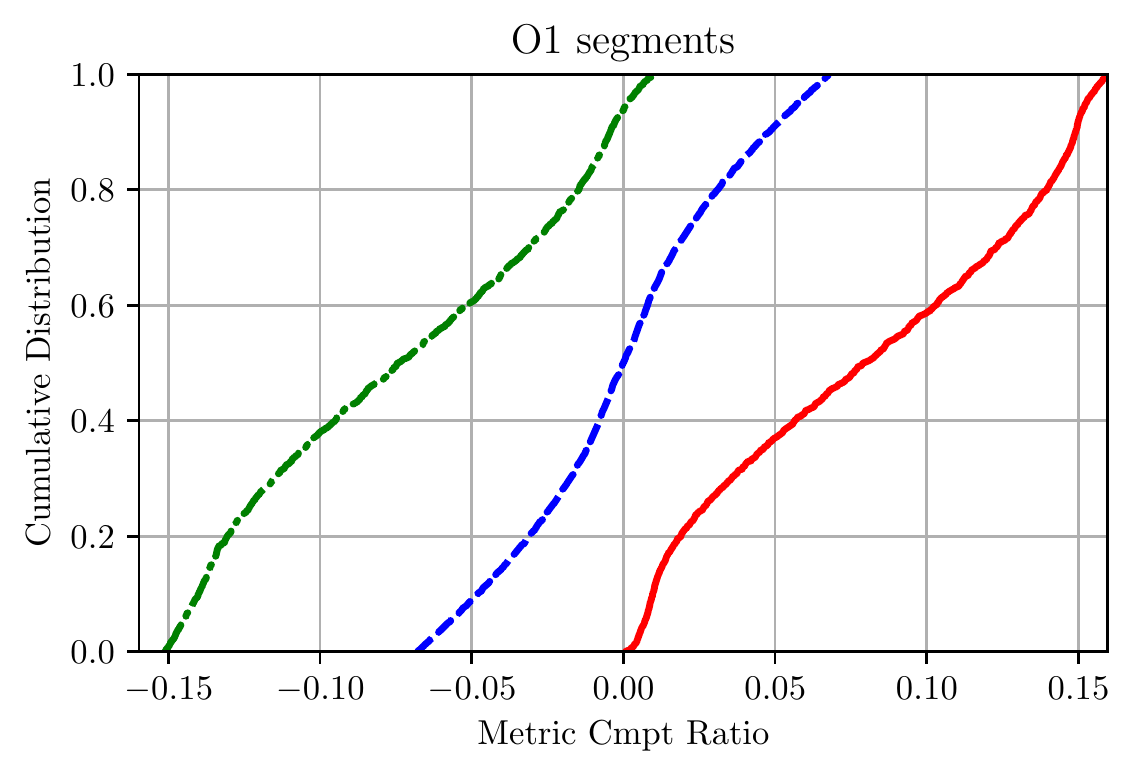}
\caption{Plots of the metric element ratios $K/I$, $L/I$, and
  $\sqrt{K^2+L^2}/I$ versus sky position of targeted source, along
  with cumulative probability distributions of these ratios, assuming a
  randomly chosen sky location, for an observation corresponding to the
  data segments (H1 and L1) from Advanced LIGO's first observing run (O1).
}
\label{f:O1_sky}
\end{center}
\end{figure}
We see that the ratios $K/I$ and $L/I$, plotted in \figref{f:O1_sky},
are small enough that a Taylor expansion should be promising.

\begin{figure}
\begin{center}
\includegraphics[width=0.48\textwidth]{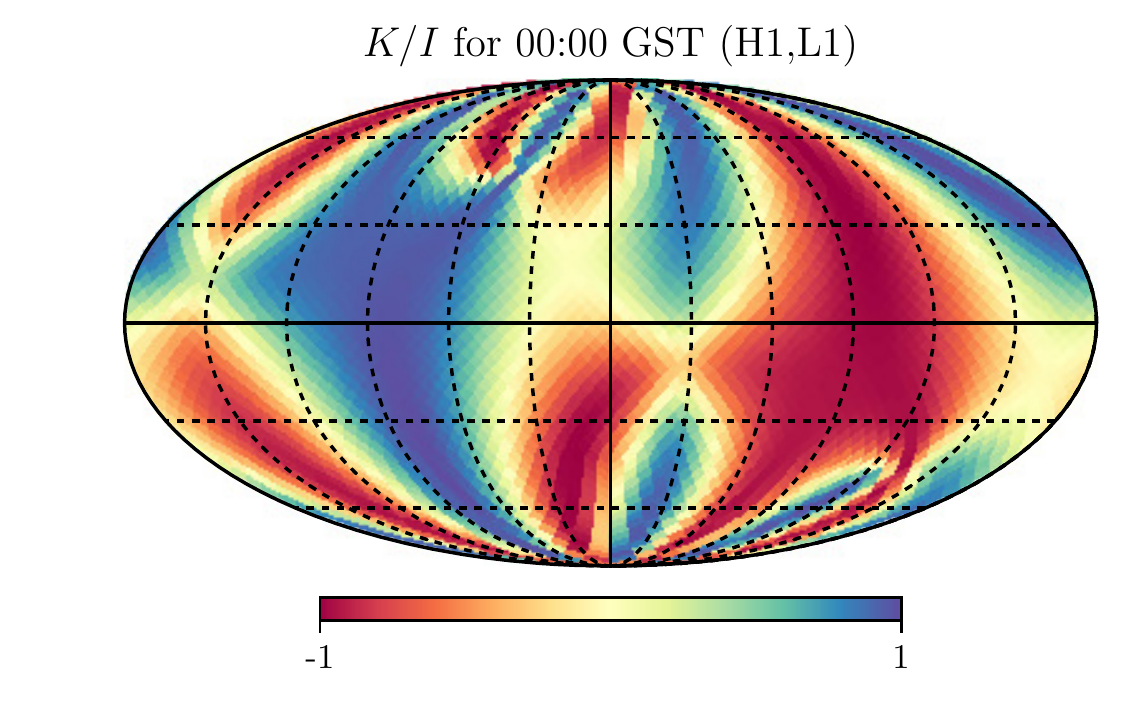}
\includegraphics[width=0.48\textwidth]{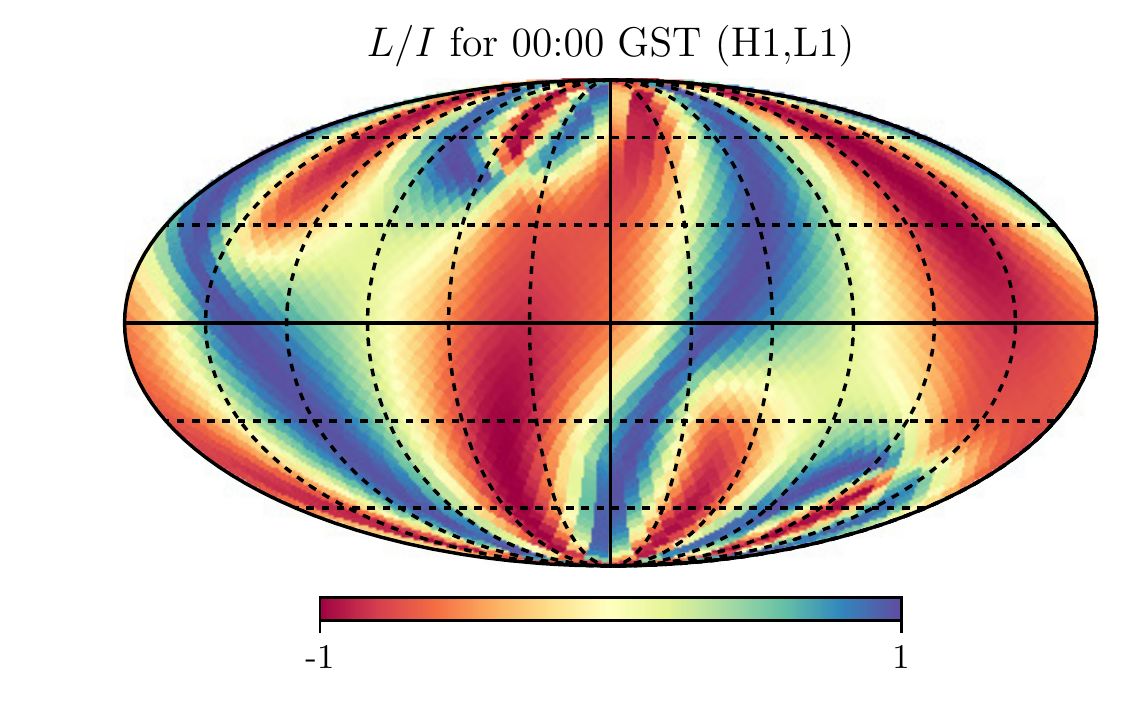}
\includegraphics[width=0.48\textwidth]{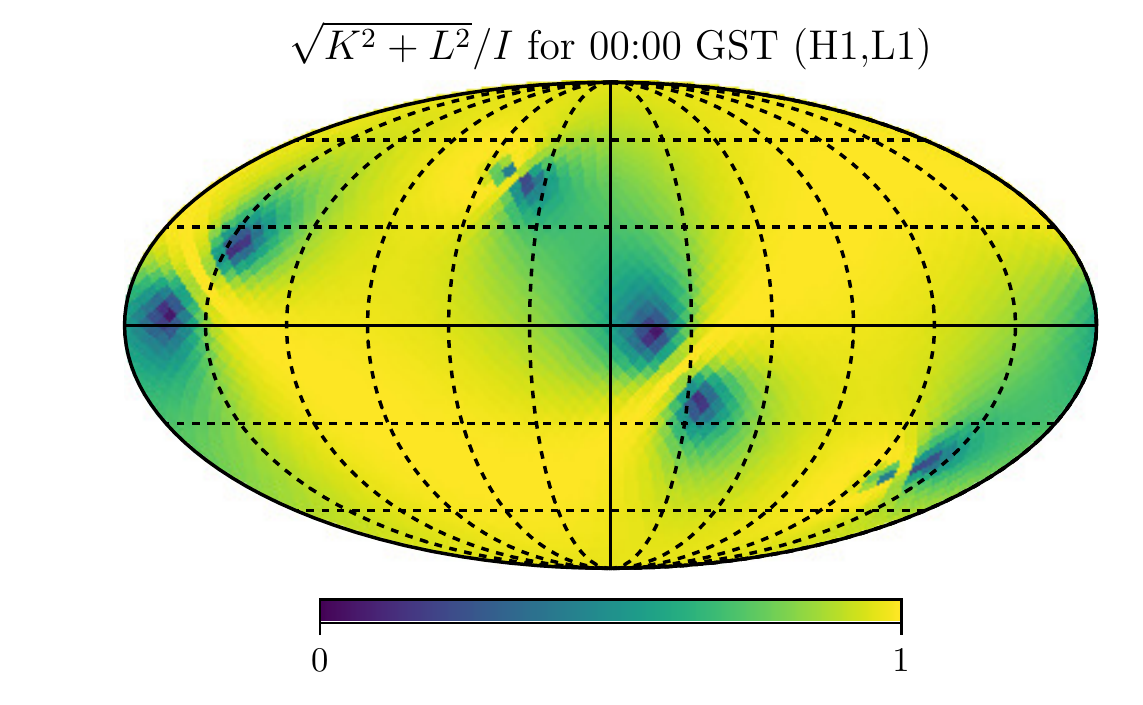}
\includegraphics[width=0.48\textwidth]{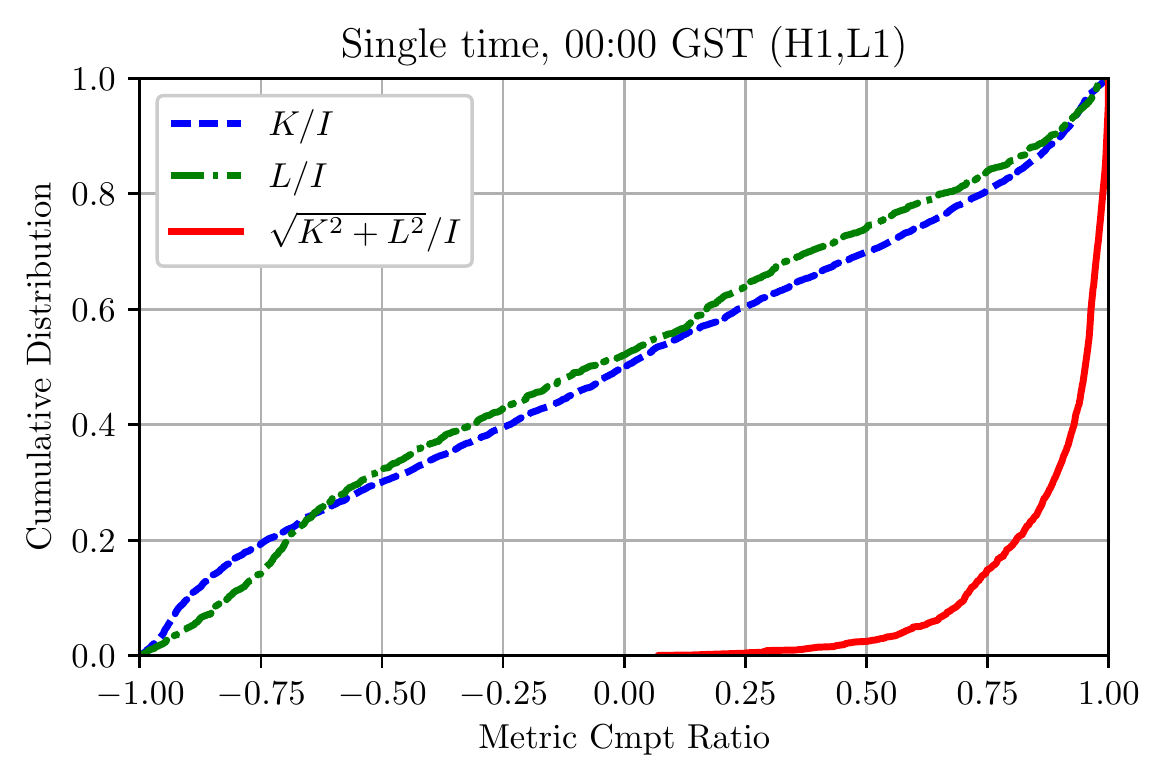}
\caption{Plots of the metric element ratios $K/I$, $L/I$, and
  $\sqrt{K^2+L^2}/I$ versus sky position of targeted source, along
  with cumulative probability distributions of these ratios, assuming a
  randomly chosen sky location, for a brief observation at Greenwich
  sidereal time 00:00.}
\label{f:GST0_sky}
\end{center}
\end{figure}

As a worst-case example (and an illustration of why this approximation
is better suited to long continuous-wave observations than to
transients), in \figref{f:GST0_sky}, we show the relevant metric
component ratios for an observation at a single time, assumed to
correspond to sidereal time 00:00 at the prime meridian.  We see that
in this case, the bound $\sqrt{K^2+L^2}\le I$ is nearly saturated for
much of the sky.

\section{Derivation of Taylor Expansion}
\label{app:taylor}

Here we collect the detailed derivation of the Taylor-expanding
$\B$-statistic.

In terms of the polar representation,
\begin{multline}
  \A^{\mudot} \M_{\mudot\nudot} \MLE{\A}^{\nudot}
  = I\AR\MLEAR\cos(\tR-\MLEtR)
  + J\AL\MLEAL\cos(\tL-\MLEtL)
  \\
  + \AR\MLEAL
  \left[
    K\sin(\tR-\MLEtL) + L\cos(\tR-\MLEtL)
  \right]
  + \AL\MLEAR
  \left[
    - K\sin(\tL-\MLEtR) + L\cos(\tL-\MLEtR)
  \right]
\end{multline}
and [see eqn (5.10) of \cite{WPCW}]
\begin{equation}
  \label{e:AMA}
  \A^{\mudot} \M_{\mudot\nudot} \A^{\nudot}
  = I\AR^2
  + J\AL^2
  + 2\AR\AL
  \left[
    K\sin(\tR-\tL) + L\cos(\tR-\tL)
  \right]
\end{equation}
so that
\begin{equation}
  \begin{split}
    \Lambda(\A;\detV{x})
    =&\ I
    \left(
      -\frac{1}{2}\AR^2 + \AR\MLEAR\cos(\tR-\MLEtR)
    \right)
    + J
    \left(
      -\frac{1}{2}\AL^2 + \AL\MLEAL\cos(\tL-\MLEtL)
    \right)
    \\
    &+ K
    \left(
      -\AR\AL\sin(\tR-\tL)
      +\AR\MLEAL\sin(\tR-\MLEtL)
      -\MLEAR\AL\sin(\tL-\MLEtR)
    \right)
    \\
    &+ L
    \left(
      -\AR\AL\cos(\tR-\tL)
      +\AR\MLEAL\cos(\tR-\MLEtL)
      +\MLEAR\AL\cos(\tL-\MLEtR)
    \right)
    \\
    =&\ \LambdaR(\AR,\tR;\MLEAR,\MLEtR) + \LambdaL(\AL,\tL;\MLEAL,\MLEtL)
    \\
    &+
    \left[
      K\sin(\MLEtR-\MLEtL) + L\cos(\MLEtR-\MLEtL)
    \right]
    \\
    &\phantom{+}\times
    \Bigl[
    \AR\AL
    \left(
      - \cos(\tR-\MLEtR)\cos(\tR-\MLEtR)
      + \sin(\tR-\MLEtR)\sin(\tL-\MLEtL)
    \right)
    \\
    &\phantom{+\times\Bigl[}
    +\AR\MLEAL\cos(\tR-\MLEtR)
    +\MLEAR\AL\cos(\tL-\MLEtL)
    \Bigr]
    \\
    &+
    \left[
      K\cos(\MLEtR-\MLEtL) + L\sin(\MLEtR-\MLEtL)
    \right]
    \\
    &\phantom{+}\times
    \biggl[
    \AR\AL
    \left(
      \cos(\tR-\MLEtR)\sin(\tL-\MLEtL)
      + \sin(\tR-\MLEtR)\cos(\tL-\MLEtL)
    \right)
    \\
    &\phantom{+\times\Bigl[}
    +\AR\MLEAL\sin(\tR-\MLEtR)
    +\MLEAR\AL\sin(\tL-\MLEtL)
    \biggr]
  \end{split}
\end{equation}
  The likelihood ratio can be expanded, to
first order, as
\begin{equation}
  e^{\Lambda(\{\A^{\mudot}\};\detV{x})}
  = e^{
    \LambdaR(\AR,\tR;\MLEAR,\MLEtR) + \LambdaL(\AL,\tL;\MLEAL,\MLEtL)
    + \Lambda_1(\A;\MLE{\A})
  }
  \approx e^{\LambdaR(\AR,\tR;\MLEAR,\MLEtR)}e^{\LambdaL(\AL,\tL;\MLEAL,\MLEtL)}
  \left(
    1 + \Lambda_1(\A;\MLE{\A})
  \right)
\end{equation}
In this form, we can factor the integrals in each of the terms; they
all reduce to one of three forms:
\begin{subequations}
  \begin{gather}
    \label{eq:JAint0}
    \int_{0}^{2\pi}\int_{0}^{\infty}
    e^{-\frac{I}{2}A^2+IA\MLE{A}\cos(\phi-\MLE{\phi})} \frac{dA\,d\phi}{\sqrt{A}}
    = 2\pi\int_{0}^{\infty} e^{-\frac{I}{2}A^2}I_0(IA\MLE{A}) \frac{dA}{\sqrt{A}}
    \\
    \label{eq:JAint1}
    \int_{0}^{2\pi}\int_{0}^{\infty}
    e^{-\frac{I}{2}A^2+IA\MLE{A}\cos(\phi-\MLE{\phi})}\cos(\phi-\MLE{\phi})
    \,\sqrt{A}\,dA\,d\phi
    = 2\pi\int_{0}^{\infty} e^{-\frac{I}{2}A^2}I_1(IA\MLE{A}) \sqrt{A}\,dA
    \\
    \int_{0}^{2\pi}\int_{0}^{\infty}
    e^{-\frac{I}{2}A^2+IA\MLE{A}\cos(\phi-\MLE{\phi})}\sin(\phi-\MLE{\phi})
    \,\sqrt{A}\,dA\,d\phi
    = 0
  \end{gather}
\end{subequations}
where $I_n(x)=i^{-n}J_n(ix)$ is the modified Bessel function, and we
have used the Jacobi-Anger
expansion\cite{abramowitz64:_handb_mathem_funct}, which tells us that
\begin{equation}
  e^{IA\MLE{A}\cos(\phi-\MLE{\phi})}
  = I_0(IA\MLE{A}) + 2\sum_{n=1}^{\infty} I_n(IA\MLE{A})\cos(n[\phi-\MLE{\phi}])
  \ .
\end{equation}
Both of the remaining integrals can be done using equation (11.4.28)
of \cite{abramowitz64:_handb_mathem_funct}, which says, in terms of
the modified Bessel function, that, when $\Real(\nu+\mu)>0$ and
$\Real(a^2)>0$,
\begin{equation}
  \label{e:ASIint}
  \int_{0}^{\infty} e^{-a^2t^2}\,t^{\mu-1} I_{\nu}(bt)\,dt
  = \frac{\Gamma\left(\frac{\nu+\mu}{2}\right)\left(\frac{b}{2a}\right)^\nu}
  {2a^\mu\Gamma(\nu+1)}
  \ {}_1F_1\left(\frac{\nu+\mu}{2},\nu+1,\frac{b^2}{4a^2}\right)
\end{equation}
where ${}_1F_1(a,b,z)=M(a,b,z)$ is the confluent hypergeometric
function.  We apply this with $a^2=I/2$, $b=I\MLE{A}$, $\mu=1/2$ and
$3/2$, and $\nu=0$ and $1$, respectively, in \eqref{eq:JAint0} and
\eqref{eq:JAint1}, to get
\begin{subequations}
  \label{e:1F1int}
  \begin{gather}
    \int_{0}^{2\pi}\int_{0}^{\infty}
    e^{-\frac{I}{2}A^2+IA\MLE{A}\cos(\phi-\MLE{\phi})} \frac{dA\,d\phi}{\sqrt{A}}
    = 2\pi\frac{\Gamma\left(\frac{1}{4}\right)}
    {2^{3/4}I^{1/4}}
    \ {}_1F_1\left(\frac{1}{4},1,\frac{I\MLE{A}^2}{2}\right)
    \\
    \int_{0}^{2\pi}\int_{0}^{\infty}
    e^{-\frac{I}{2}A^2+IA\MLE{A}\cos(\phi-\MLE{\phi})}\cos(\phi-\MLE{\phi})
    \,\sqrt{A}\,dA\,d\phi
    = 2\pi\frac{\Gamma\left(\frac{5}{4}\right)\MLE{A}}
    {2^{3/4}I^{1/4}}
    \ {}_1F_1\left(\frac{5}{4},2,\frac{I\MLE{A}^2}{2}\right)
  \end{gather}
\end{subequations}
We can use these to evaluate the integral for the $\B$-statistic
\eqref{e:Bstatexplicit} as
\begin{multline}
  \B(\detV{x})
  \approx
  \frac{A}{8\pi^2}
  \left(2\pi\frac{\Gamma\left(\frac{1}{4}\right)}{2^{3/4}I^{1/4}}\right)
  \left(2\pi\frac{\Gamma\left(\frac{1}{4}\right)}{2^{3/4}J^{1/4}}\right)
  {}_1F_1\left(\frac{1}{4},1,\frac{I\MLEAR^2}{2}\right)
  {}_1F_1\left(\frac{1}{4},1,\frac{J\MLEAL^2}{2}\right)
  \\
  \times
  \left\{
    1 +
    \left[
      K\sin(\MLEtR-\MLEtL) + L\cos(\MLEtR-\MLEtL)
    \right]
    \MLEAR\MLEAL
    \vphantom{\frac{{}_1F_1\left(\frac{5}{4},2,\frac{J\MLEAL^2}{2}\right)}{{}_1F_1\left(\frac{1}{4},1,\frac{J\MLEAL^2}{2}\right)}}
  \right.
  \\
  \left.
    \times
    \left[
      \frac{1}{4}
      \left(
        \frac{{}_1F_1\left(\frac{5}{4},2,\frac{I\MLEAR^2}{2}\right)}{{}_1F_1\left(\frac{1}{4},1,\frac{I\MLEAR^2}{2}\right)}
      \right)
      + \frac{1}{4}
      \left(
        \frac{{}_1F_1\left(\frac{5}{4},2,\frac{J\MLEAL^2}{2}\right)}{{}_1F_1\left(\frac{1}{4},1,\frac{J\MLEAL^2}{2}\right)}
      \right)
      -
      \frac{1}{16}
      \left(
        \frac{{}_1F_1\left(\frac{5}{4},2,\frac{I\MLEAR^2}{2}\right)}{{}_1F_1\left(\frac{1}{4},1,\frac{I\MLEAR^2}{2}\right)}
      \right)
      \left(
        \frac{{}_1F_1\left(\frac{5}{4},2,\frac{J\MLEAL^2}{2}\right)}{{}_1F_1\left(\frac{1}{4},1,\frac{J\MLEAL^2}{2}\right)}
      \right)
    \right]
  \right\}
\end{multline}

\section{Recovery of $\F$-Statistic}

Our method expands the $\B$-statistic to first order in the metric
components $K$ and $L$.  It has been shown in \cite{PK}
that the Bayes factor constructed with a prior uniform in the
$\{\A^{\mudot}\}$ is equivalent to the $\F$-statistic, which we note
in \eqref{e:Fstat} has only zeroth- and first-order terms in these
quantities.  This means that applying the Taylor-expansion method with
this prior should reproduce the \emph{exact} $\F$-statistic.

If we replace the isotropic prior \eqref{e:Bstatprior} with a uniform
prior
$\pdf(\A^{\onedot},\A^{\twodot},\A^{\tredot},\A^{\fordot}|\Hf)=C$,
the $\B$-statistic integral \eqref{e:Bstatexplicit} becomes
\begin{equation}
  \begin{split}
    \B(\detV{x})
    &=
    C
    \int_{-\infty}^{\infty}
    \int_{-\infty}^{\infty}
    \int_{-\infty}^{\infty}
    \int_{-\infty}^{\infty}
    e^{\Lambda(\{\A^{\mudot}\};\detV{x})}
    \,d\A^{\onedot}
    \,d\A^{\twodot}
    \,d\A^{\tredot}
    \,d\A^{\fordot}
    \\
    &=
    C
    \int_{0}^{2\pi}
    \int_{0}^{2\pi}
    \int_{0}^{\infty}
    \int_{0}^{\infty}
    e^{\Lambda(\{\A^{\mudot}\};\detV{x})}
    \AR\AL\,d\AR\,d\AL\,d\tR\,d\tL
  \end{split}
\end{equation}
The Taylor expansion of the likelihood, and the angular integrals,
proceed as in \appref{app:taylor}, and the only difference is that the
two principal integrals
\eqref{eq:JAint0} and \eqref{eq:JAint1}, become
\begin{subequations}
  \begin{gather}
    \int_{0}^{2\pi}\int_{0}^{\infty}
    e^{-\frac{I}{2}A^2+IA\MLE{A}\cos(\phi-\MLE{\phi})} \,A\,dA\,d\phi
    = 2\pi\int_{0}^{\infty} e^{-\frac{I}{2}A^2}I_0(IA\MLE{A}) \,A\,dA
    \\
    \int_{0}^{2\pi}\int_{0}^{\infty}
    e^{-\frac{I}{2}A^2+IA\MLE{A}\cos(\phi-\MLE{\phi})}\cos(\phi-\MLE{\phi})
    \,A^2\,dA\,d\phi
    = 2\pi\int_{0}^{\infty} e^{-\frac{I}{2}A^2}I_1(IA\MLE{A}) \,A^2\,dA
  \end{gather}
\end{subequations}
Using \eqref{e:ASIint} with $a^2=I/2$, $b=I\MLE{A}$, $\mu=2$ and $3$,
and $\nu=0$ and $1$, respectively, we find, in place of
\eqref{e:1F1int},
\begin{subequations}
  \label{e:1F1int}
  \begin{gather}
    \int_{0}^{2\pi}\int_{0}^{\infty}
    e^{-\frac{I}{2}A^2+IA\MLE{A}\cos(\phi-\MLE{\phi})} \,A\,dA\,d\phi
    = \frac{2\pi}{I}
    \ {}_1F_1\left(1,1,\frac{I\MLE{A}^2}{2}\right)
    = \frac{2\pi}{I}e^{\frac{I\MLE{A}^2}{2}}
    \\
    \int_{0}^{2\pi}\int_{0}^{\infty}
    e^{-\frac{I}{2}A^2+IA\MLE{A}\cos(\phi-\MLE{\phi})}\cos(\phi-\MLE{\phi})
    \,A^2\,dA\,d\phi
    = \frac{2\pi\MLE{A}}{I}
    \ {}_1F_1\left(2,2,\frac{I\MLE{A}^2}{2}\right)
    = \frac{2\pi\MLE{A}}{I}e^{\frac{I\MLE{A}^2}{2}}
  \end{gather}
\end{subequations}
where we have used (13.6.12) of
\cite{abramowitz64:_handb_mathem_funct}, which states that
${}_1F_1(a,a,z)=e^z$.  This then gives a statistic of
\begin{equation}
  \B(\detV{x})
  \approx
  C
  \left(\frac{2\pi}{I}\right)
  \left(\frac{2\pi}{J}\right)
  e^{\frac{I\MLEAR^2+J\MLEAL^2}{2}}
  \left\{
    1 +
    \left[
      K\sin(\MLEtR-\MLEtL) + L\cos(\MLEtR-\MLEtL)
    \right]
    \MLEAR\MLEAL
    (1+1-1)
  \right\}
\end{equation}
So that, to first order in $K$ and $L$,
\begin{equation}
  \ln\frac{\B(\detV{x})}{\B(\detV{0})}
  \approx
  \frac{I\MLEAR^2}{2}+\frac{J\MLEAL^2}{2}
  +
  \left[
    K\sin(\MLEtR-\MLEtL) + L\cos(\MLEtR-\MLEtL)
  \right]
  \MLEAR\MLEAL
\end{equation}
Which is indeed the form given in \eqref{e:Fstat} for the exact
$\F$-statistic.

\section{Relationship to High-SNR Approximation}

\label{app:DKW}

Recent work\cite{DKW} by Dhurandhar, Krishnan and Willis (hereafter
DKW) contains a different approximate expression for the
$\B$-statistic, derived in the limit of high signal-to-noise ratio,
but without assumptions on the form of the metric.  In their notation,
the approximate form is written [\cite{DKW} equation (104)]
\begin{equation}
  \B(x) \approx \left(\frac{\pi^2}{2(\zeta^2 - k^2)}\right)
  \left[
    \frac{
      e^{
        \frac{1}{2}\widehat{\mathbf{B}}^{\dagger}
        \mathbf{N}\widehat{\mathbf{B}}}
    }
    {(|\widehat{\B}_1||\widehat{\B}_2|)^{\frac{3}{2}}}
  \right]
\label{eq:DKW_non_zero}
\end{equation}
To make contact with our results, we collect here the conversion
between DKW's notation and ours.  Their metric elements are
$\zeta=I=J$ (they limit attention to the long-wavelength limit) and
$\kappa=L+iK$, with $k=\abs{\kappa}=\sqrt{K^2+L^2}$.  They define
complex amplitudes
\begin{subequations}
  \label{eq:complex_phys}
  \begin{align}
    \B_1 &= h_0 e^{-2i\phi_0}\frac{(1+\cosi)^2}{4}e^{-2i\psi}
           = \AR e^{-\frac{i}{2}(3\tR + \tL)} = \B_4^*
    \\
    \B_2 &= h_0 e^{-2i\phi_0}\frac{(1-\cosi)^2}{4}e^{2i\psi}
           = \AL e^{-\frac{i}{2}(\tR + 3\tL)} = \B_3^*
  \end{align}
\end{subequations}
and a complex metric
\begin{equation}
  \label{eq:Ndot}
  \textbf{N} \equiv\{N_{\mu\nu}\}
  = \frac{1}{2}
  \begin{pmatrix}
    \zeta &  \kappa^* &  0 & 0 \\
    \kappa &  \zeta &  0 &  0 \\
    0 &  0 &  \zeta &  \kappa^* \\
    0 &  0 &  \kappa &  \zeta
  \end{pmatrix}
\end{equation}
from which we see
\begin{equation}
  \begin{split}
    \frac{1}{2}\widehat{\mathbf{B}}^{\dagger}
    \mathbf{N}\widehat{\mathbf{B}}
    &= \Real\left(
      \MLEAR e^{\frac{i}{2}(3\MLEtR + \MLEtL)}
      [L - iK]
      \MLEAL e^{-\frac{i}{2}(\MLEtR + 3\MLEtL)}
    \right)
    \\
    &= \frac{1}{2}I\MLEAR^2 + \frac{1}{2}J\MLEAL^2
    + \MLEAR\MLEAL
    \left[
      K\sin(\MLEtR-\MLEtL) + L\cos(\MLEtR-\MLEtL)
    \right]
    = \F
  \end{split}
\end{equation}
and therefore their approximation can be written
\begin{equation}
  \B(x) \approx \left(\frac{\pi^2}{2(IJ-K^2-L^2)}\right)
  \left[
    \frac{e^{\F}}{(\MLEAR\MLEAL)^{\frac{3}{2}}}
  \right]
\end{equation}
Written in this form, we see that the result is the same as equation
(5.37) of \cite{WPCW}.  The difference between this and
\eqref{e:Bapproxasymptotic} is the normalization constant.  (Note that
DKW use an improper prior equivalent to $A=2\pi^2$.)

\section*{References}

\providecommand{\newblock}{}

\end{document}